\documentclass[12pt,a4paper]{article}
\usepackage{a4wide}
\usepackage{amsfonts}
\begin{document}

\def\be{\begin{equation}}
\def\ee{\end{equation}}

\def\CG{{\cal G}}
\def\CC{{\cal C}}
\def\CL{{\cal L}}
\def\CE{{\cal E}}
\def\CD{{\cal D}} 
\def\CH{{\cal H}} 
\def\CP{{\cal P}}
\def\CS{{\cal S}} 
\def\CO{{\cal O}}

\newcommand{\dId}{\mathrm{1\mkern-4.3mu I}}
\newcommand{\dR}{{\mathbb R}}
\newcommand{\dN}{{\mathbb N}}
\newcommand{\dZ}{{\mathbb Z}}
\newcommand{\dC}{{\mathbb C}}
\newcommand{\Diff}{\mbox{\rm Diff}}
\newcommand{\diff}{\mbox{\rm diff}}
\newcommand{\semidir}{\mathrm{%
\times\mkern-3.3mu\protect\rule[0.04ex]{0.04em}{1.05ex}\mkern3.3mu\mbox{}}}
\newcommand{\zq}{\overline{z}}

\newtheorem{theo}{Theorem} \newtheorem{lemma}[theo]{Lemma}

\newcommand{\proofend}{\raisebox{1.3mm}{%
\fbox{\begin{minipage}[b][0cm][b]{0cm}\end{minipage}}}}
\newenvironment{proof}[1]
{{\noindent\it Proof #1:}
}{\mbox{}\hfill \proofend\\\mbox{}}

{\renewcommand{\thefootnote}{\fnsymbol{footnote}}
\hfill  PITHA -- 99/25\\
\medskip
\hfill ESI -- 742\\
\medskip
\begin{center}
{\LARGE Group Theoretical Quantization\\
  and the Example of a Phase Space $S^1\times\dR^+$}\\ 
\vspace{1.5em}
Martin Bojowald\footnote{e-mail: {\tt bojowald@physik.rwth-aachen.de}}
and Thomas
Strobl\footnote{e-mail: {\tt tstrobl@physik.rwth-aachen.de}}\\
Institut f\"ur Theoretische Physik, RWTH-Aachen\\
D--52056 Aachen, Germany\\
\vspace{1.5em}
\begin{abstract}
  The group theoretical quantization scheme is reconsidered by means
  of elementary systems.  Already the quantization of a particle on a
  circle shows that the standard procedure has to be supplemented by
  an additional condition on the admissibility of group actions. A
  systematic strategy for finding admissible group actions for
  particular subbundles of cotangent spaces is developed,
  two--dimensional prototypes of which are $T^*\dR^+$ and
  $\CS=S^1\times\dR^+$ (interpreted as restrictions of $T^*\dR$ and
  $T^*S^1$ to positive coordinate and momentum, respectively).  In
  this framework (and under an additional, natural condition) an
  $SO^\uparrow(1,2)$--action on $\CS$ results as the {\em unique\/}
  admissible group action.

For symplectic manifolds which are (specific) parts of phase spaces
with known quantum theory a simple ``projection method'' of
quantization is formulated. For $T^*\dR^+$ and $\CS$ equivalent
results to those of more established (but more involved) quantization
schemes are obtained. The approach may be of interest, e.g., in
attempts to quantize gravity theories where demanding nondegenerate
metrics of a fixed signature imposes similar constraints.
\end{abstract}
\end{center}}
\setcounter{footnote}{0}

\section{Introduction}

To quantize a classical phase space there are techniques generalizing
the standard quantization method which is only applicable to simple
cotangent bundles. Most prominent are geometric \cite{Woodhouse} and
group theoretical quantization \cite{Isham,Gui}. But since each method
of quantization has its own advantages and disadvantages and gives
rise to certain types of ambiguities, none of them can be regarded as
a final and unique route to a quantum theory. Usually, such a scheme
is developed on simple examples so as to reproduce standard results.
Studying more complicated systems then can lead to the necessity of
further specifications which are necessary to exclude unphysical
results.

The present paper is divided into two main parts. In the first one,
Sec.~\ref{s:quant}, we review the group theoretical quantization (following
Ref.~\cite{Isham}) focussing on some points which, in our opinion,
deserve further study.  In this scheme one studies the irreducible
unitary representations of a group which has a transitive, almost
effective Hamiltonian action with a momentum map on the phase space
(keeping only those representations which are physically acceptable).
The main ingredients of this method and its application to the
simplest examples (phase space $T^* \dR$, $T^* \dR^+$, and $T^* S^1$)
are recalled here. In the course of this review we will see the
necessity of a global generating principle for phase space functions
as opposed to a local one, which is already implied by transitivity of
the group action. Therefore, we have to supplement the rules of group
theoretical quantization as outlined in Ref.~\cite{Isham} by a further
condition on the allowed group actions in order to recover the results
of standard quantizations of $T^*S^1$.

Thereafter a general strategy for finding an appropriate group action
on particular subbundles of cotangent bundles is discussed. This
extends considerations presented in Ref.~\cite{Isham} for actions on
cotangent bundles.

Sec.\ \ref{s:quant} is concluded with a proposal 
(``projection quantization'') 
for the quantization of certain submanifolds $\CP$ of phase spaces
$\widetilde \CP$, where quantum realizations of $\widetilde \CP$ are
known.

$T^* \dR^+$ may be used as illustrating example. As a restriction of
$T^* \dR$ it fits into the framework of both of the final two
subsections of Sec.\ \ref{s:quant}. Standard results are reproduced in
this case.

In Sec.~\ref{s:S} we will consider the phase space ${\cal
  S}:=S^1\times\dR^+:=T^*S^1|_{p>0}$ defined as the restriction of the
cotangent bundle of $S^1$ to positive momenta, which is maybe the
simplest example for a phase space which is not symplectomorphic to a
cotangent bundle.  It will be found that $SO^{\uparrow}(1,2)$, the
identity component of $SO(1,2)$, provides an appropriate action on
${\cal S}$ for applying group theoretical quantization.  To find this
group action we exploit the fact that ${\cal S}$ is a subbundle of
$T^*S^1$ for which we can apply the methods of Sec.~\ref{s:quant}
described above.  It will be seen that the canonical lift of an
$SO^{\uparrow}(1,2)$ subgroup of the diffeomorphism group of $S^1$ to
$T^* S^1$ provides a {\em transitive}\/ and effective Hamiltonian
action on the {\em half}\/ cylinder $\CS=S^1 \times \dR^+$. Under a
further condition we will be able to show that any such subgroup of
the lift of the diffeomorphisms is necessarily isomorphic to a
covering group of $SO^{\uparrow}(1,2)$.  All effective actions of
proper covering groups of $SO^{\uparrow}(1,2)$, which would be allowed
according to the commonly used rules of group theoretical
quantization, will be seen to be excluded by the additional condition
mentioned above.

Applying standard knowledge on the unitary irreducible representations
(IRREPs) of $SO^{\uparrow}(1,2)$ and its covering groups leads to
possible quantum realizations of the system under consideration.
Actually, as any representation of a group is also a representation of
a covering group of that group, it is sufficient to analyze the
unitary IRREPs of the universal covering group
$\widetilde{SO}\mbox{}^{\uparrow}(1,2)$ of $SO^{\uparrow}(1,2)$ to
obtain the most general possible quantum theory. This in turn amounts
to an analysis of the unitary IRREPs of the Lie algebra of $so(1,2)
\sim su(1,1) \sim sl(2,\dR)$. Selecting appropriate representations
which fulfill the relation $p>0$ at the quantum level will complete
the group theoretical quantization of $\CS$, leading to a
one--parameter family of inequivalent quantizations given by the
positive discrete series $D^k$ of
$SO^{\uparrow}(1,2)$--representations, which is labeled by a parameter
$k\in\dR^+$.

These considerations are supplementary to those in
Ref.~\cite{Loll2,Schramm}, 
where the group theoretical quantization of $\CS$
has been carried out already. Other recent related work is 
Ref.~\cite{MRT,Louko,Trunk}, where the quantization of a system is
discussed the reduced phase space of which (or, rather, its regular
part) turns out to be the four--fold copy of our phase space ${\cal
  S}$.

By definition, $\CS$ is the restriction of the cotangent bundle
$T^*S^1$ to positive momenta. The above mentioned projection quantization 
may therefore be used as an alternative (and simple) route to the
quantization of $\CS$.  Equivalence of this quantization with the
group theoretical one restricts the parameter $k$ to the interval
$0<k\leq 1$ (due to a maximality condition in the projection quantization). 
This restricted range for $k$ coincides also with what one expects on
general grounds \cite{Woodhouse} for a phase space with fundamental
group $\pi_1=\dZ$ ($\theta$-angle). In the group theoretical
quantization, however, all positive values of $k$, labeling the
inequivalent $so(1,2)$--representations of the positive discrete
series, come out on an equal footing. (Here, $k$ can be restricted to
the interval $0<k\leq 1$ by regarding representations with $k>1$ as
``unphysical'' --- it is not unusual that not all possible unitary
IRREPs are physically acceptable and thus taken into account. However,
here this exclusion cannot be done ``intrinsically'' such as, e.g., in
terms of an operator condition.)  On the other hand, relaxing the
maximality condition in the projection quantization, 
all the values of $k
\in \dR^+$ can be realized also there. This is seen to lead to an
apparently novel realization of the positive discrete series in terms of
functions over $S^1$.

In the context of projection quantization, 
our phase space can also be
viewed as a toy model for imposing similar constraints, e.g., the
constraint $\det e>0$ in a dreibein formulation of general relativity.
This analogy, and some of its limitations, are discussed briefly
in Subsec.\ \ref{s:restriction}.

A further application of our considerations to a gravitational problem
can be found in Ref.~\cite{Schramm}, where it is shown that a suitable
periodic identification of the reduced phase space of Schwarzschild
black holes in an arbitrary spacetime dimension yields the phase space
${\cal S}$.

\section{Quantization}
\label{s:quant}

We begin this section by briefly recapitulating 
the group theoretical quantization scheme by means of some elementary
systems. To reobtain the standard results for the quantization of
$T^*S^1$, we will find it necessary to reconsider the generating
principle: Already this simple example illustrates the necessity to
require that {\em any\/} function on phase space can be generated
(globally) by means of the fundamental observables obtained from the
momentum map of the group action (``strong generating principle''). As
shown in Subsec.\ \ref{Generating}, in many cases (such as, e.g., when
the group $\CG$ under discussion is semisimple) it turns out to be
sufficient to simply consider the center of the group $\CG$ (or the
center of a group $G$ closely related to $\CG$, cf Lemma 1 below) as
to the effect of excluding a candidate $\CG$--action (instead of
explicitly checking the strong generating principle for the respective
set of fundamental observables).

In Subsec.\ \ref{s:sub} we collect some of the remarks of
Ref.~\cite{Isham} on group actions on general cotangent bundles
(related to lifts of the diffeomorphism group of the base manifold).
The situation will be found to simplify considerably when certain
subbundles are considered, which, in the two--dimensional case, are
nothing but $T^*\dR^+$, $\CS$, or disjoint unions of these two. This
leads to a general strategy of finding admissible group actions on
such subbundles, which is then subsequently illustrated for both of
the two--dimensional cases.

Finally, a projection method of quantization 
is introduced in Subsec.\ \ref{s:restriction} which is also applicable
to these two--dimensional examples. For higher--dimensional phase
spaces it is not applicable to the subbundles considered in Subsec.\ 
\ref{s:sub}, while, on the other hand, its range of applicability is
much wider.

\subsection{Review of the group theoretical quantization scheme}
\label{s:group}

The Heisenberg commutation relations \be [q^i,p_j] = i
\hbar \delta^i_j \qquad \mbox{(all other commutators vanishing)}
\label{Heisenberg} \ee are at the heart of many introductory textbooks
on quantum mechanics.  Mathematically they are, however, not an
adequate {\em starting point}\/ for quantization. Firstly, these
relations can certainly be valid only on a dense subspace of the full
Hilbert space.  But even worse, there exist many inequivalent (and
unphysical) representations of these relations on a dense subspace (cf
Ref.~\cite{Thirr3Aufl2}, p.\ 88 for a simple example), which in part
is connected to the fact that the commutation relations take into
account only local information about the phase space (cf the example
in Ref.~\cite{Isham}, p.\ 1131).  The ``exponentiated'' Heisenberg
relations, defining the Weyl algebra, on the other hand, {\em are}\/ a
good starting point for an algebraic approach to quantization.  With
$U(a) := \exp(-ia^j p_j)$ and $V(b) := \exp(-ib_jq^j)$, where $a,b \in
\dR^n$ (up to respective units), the Weyl algebra has the form \be
U(a)U(a')=U(a+a') \, , \; V(b)V(b')=V(b+b') \, , \; U(a) V(b) = V(b)
U(a) e^{i\hbar a^jb_j} \, .
\label{Weyl} \ee  It is a mathematical fact that (for finite $n$ and
for fixed $\hbar$) the irreducible, strongly continuous
representations of the Weyl algebra are {\em unique}\/ (up to unitary
equivalence) and equivalent to the standard representation of quantum
mechanics in a Hilbert space $L^2(\dR^n,d^nx)$ with $q^i$ being the
multiplication operator $x^i$ and $p_i = -i\hbar d/dx^i$.  (Cf e.g.\ 
Refs.~\cite{Reed,Thirr3Aufl2,Isham} for further details).

If the configuration space is no more an $\dR^n$ or the phase space
even no more a cotangent bundle, quantization is no more that unique
and different generalizations or alterations of the above approach
come into question. E.g.\ on a configuration space $T^n$ ($n$-torus)
the relations (\ref{Weyl}) are required to hold for $b_i \in\dZ$ only
and the space of unitarily inequivalent representations becomes
$U(1)^n$ (corresponding to the different possibilities of (mutually
commuting, cf Ref.~\cite{Nico}) self--adjoint extensions of the
operators $p_i = -i\hbar d/dx^i$ on $[0,2\pi]^n$).  And on a
configuration space $\dR^+$ it is even no more adequate to consider
unitary representations of $\exp(i \lambda p)$ (among the other elements in
the Weyl algebra); the hermitean operator $p = -i\hbar d/dx$ has no
self--adjoint extensions on $\dR^+$, $\exp(i \lambda p)$ is a translation
operator that does not map $\dR^+$ into itself for all values of $\lambda$.

Geometric quantization is one of the most prominent attempts for a
quantization procedure applicable to more or less arbitrary phase
spaces (cf, e.g., Ref.~\cite{Woodhouse} for an introduction). 
Another method is the group theoretical quantization, which is
inspired in part by geometric quantization as well as by work of
Mackey \cite{Mackey}; cf Ref.~\cite{Isham} for a review.
Since this approach requires the phase space to be some coset space,
it has the drawback that even in some cases of finite dimensional
phase spaces one may be forced to use infinite dimensional groups and
their representation theory.  However, in many finite dimensional
examples of physical interest (cf e.g.\ Ref.~\cite{Isham}), including
the phase space $\CS$ studied in detail in Sec.\ \ref{s:S}, this is not the
case.

In the context of a standard configuration space $\dR^n$, the group
theoretical approach arises as follows: Reinterpreting the phase
factor in the Weyl algebra (\ref{Weyl}) as a central element, the
relations (\ref{Weyl}) may be understood as the multiplication law for a
$2n+1$--dimensional Lie group. Its Lie algebra is given by the
Heisenberg relations (\ref{Heisenberg}) with generators $q^i,p_j$, and
$1$, where the need for the latter generator results from the
right-hand side of the commutators, $1$ denoting a central element of
the full Lie algebra.  The study of the irreducible unitary
representations of this group (or its universal covering group, the
Heisenberg group) yields our standard quantum theory.

Let us analyse this situation more carefully so that it allows a
generalization to more general phase spaces: The operators $U$ and $V$
in Eq.~(\ref{Weyl}) are translation operators in the configuration space
and momentum space, respectively. {\em Each}\/ of these
$n$--dimensional translation groups has an analogue in the classical
phase space $\CP \equiv T^* \dR^n$, $q \to q + a$ and $p \to p + b$.
Put together, these two transformation groups form the
$2n$--dimensional {\em abelian}\/ group\footnote{This is in contrast
to putting together the operators $U$ and $V$, as seen by the last
relation in Eq.~(\ref{Weyl}); we will shortly come back to this
difference.} $\CG = (\dR^{2n},+)$, which acts transitively and
effectively on $\CP$ and leaves the symplectic form $\omega= dq^i \wedge
dp_i$ invariant, i.e.\ it is a group of canonical transformations.
({\em Transitivity}\/ means that for any two points in $\CP$ there is
a group element such that its application to one of the points yields
the other one, {\em effectiveness}\/
implies that only the identity of $\CG$ acts trivially on $\CP$).  The
action of $\CG$ is Hamiltonian, moreover, i.e.\ there exist (globally
defined) functions $F_A$ on $\CP$, such that the vector fields $V_A
\equiv \{ \cdot, F_A \}$ generate the group action (the invariance of
$\omega$ guarantees only the {\em local}\/ existence of functions $F_A$);
in the present case with group $(\dR^{2n},+)$ the respective
Hamiltonians (or observables) are $q^i$, $p_i$, $i= 1, \ldots, n$ (up
to an addition of constants, which drop out from the generating vector
fields $V_A$).

In the general case of an (effective) action of a group $\CG$ on a
phase space $\CP$, the Lie bracket of the generating vector fields
$V_A$ always mimics the Lie algebra $\CL(\CG)$ of the group $\CG$:
$[V_A, V_B]=f_{AB}^C \, V_C$, where $f_{AB}^C$ are the structure
constants of $\CL(\CG)$ (in the above case of an abelian group $G$,
$f_{AB}^C \equiv 0$).  If the action is Hamiltonian, one may conclude
from this in general only that $\{F_A,F_B \} = - f_{AB}^C \, F_C +
\kappa(A,B)$, where $\kappa$ is a constant on $\CP$. As a function on
the Lie algebra, $\kappa$ is a two--cocycle (as a consequence of the
Jacobi identity for the Poisson bracket), which changes by a
two--coboundary upon redefining the functions $F_A$ by a constant;
thus $\kappa \in H^2(\CL(\CG),\dR)$ (cf e.g.\ Ref.~\cite{Woodhouse} for
more details on this and related aspects).  In many cases the
constants $\kappa$ can be made to vanish upon an appropriate choice of
functions $F_A$ in which case the Hamiltonian action is said to allow
a {\em momentum map}. E.g., this is the case when the group $\CG$ is
semisimple, because then $H^2(\CL(\CG),\dR)$ is trivial. However, also
any Hamiltonian action of a Lie group on a {\em compact\/} phase space
has a momentum map (independently of the second cohomology of the
respective group) or, similarly, any subgroup of the lift of the
diffeomorphism group of an arbitrary configuration space.

It is obvious from the Poisson brackets $\{ q^i,p_j\}=\delta^i_j$ that
the action of $(\dR^{2n},+)$ on $T^* \dR^n$ does not allow a momentum
map (clearly the righthand side of these relations cannot be removed
by shifting $q^i$ and $p_j$ by constants). Although $T^*\dR^n$ is the
simplest choice of a phase space, from the point of view of group
theoretical quantization it is rather an involved example (due the
absence of a momentum map).  Instead of $\CG = (\dR^{2n},+)$ one is
then lead to focus on a central extension $\CE$ of this group, the Lie
algebra of which may be spanned by $F_A$ and a central element $1$,
with the Lie bracket provided by the Poisson bracket between the
corresponding functions on $\CP$. The unique simply connected choice
for this group $\CE$ is the Heisenberg group. Note that $(\dR^{2n},+)$
is {\em not}\/ a subgroup of $\CE$, any abelian subgroup having at
most dimension $n+1$; only the factor group $\CE/\dR$ with respect to
the central subgroup ${\cal N}=\dR$ yields $(\dR^{2n},+)$. Also, in
contrast to the latter group, $\CE$ does not act effectively on $\CP$
anymore (as ${\cal N}$ acts trivially on $\CP$), while it certainly still
is transitive.

There is a one--parameter family of weakly continuous
unitary\footnote{As a consequence of unitarity, weak continuity
  implies strong continuity. Alternatively, we may
require strong continuity and then find,
although only for specific cases such as  for the Heisenberg group,
that all the representations are unitary.}
IRREPs of the Heisenberg group $\CE$. This parameter stems from the
unitary representation of the central subgroup ${\cal N}=(\dR,+)$.
Following Isham, the freedom in this parameter is fixed by nature's
value of $\hbar$. This brings us back to the first paragraph of this
section with its unique quantum theory.

Preliminarily, we state the following general strategy in the group
theoretical approach to a quantum theory for a given phase space $\CP$
(cf Ref.~\cite{Isham} for further motivation and details): First find a
Hamilonian, transitive, and almost effective\footnote{I.e., there is
  only a {\em discrete}\/ set of elements of $\CG$ (which necessarily
  is an invariant subgroup) acting trivially on all points of $\CP$.}
action of a group $\CG$ on $\CP$.  {\em If}\/ this action allows a
momentum map, the next and final step is to study the weakly continuous,
unitary IRREPs of $\CG$ (discarding possibly physically unacceptable
representations). If, on the other hand, there is no momentum map for
the action of $\CG$, one again considers the one--parameter central
extension $\CE$ of $\CG$ and then studies the weakly continuous,
unitary IRREPs of $\CE$.

In general there may be different admissible groups acting on the
phase space and each of these groups may have different, inequivalent
actions. Moreover, for any group there may be various admissible
unitary representations. Some of the latter may be excluded upon
physical considerations (such as, e.g., by positivity of a classically
positive Hamiltonian), the possible ambiguity in the remaining IRREPs
being interpreted as part of the ambiguity in the transition from a
classical system to its quantum version.

Note that clearly {\em any\/} function on $T^*\dR^n$ is a function of
the elementary observables $q^i$ and $p_i$. As will be found below, a
similar requirement on the fundamental observables $F_A$ of the group
$\CG$ {\em has\/} to be asked for also in the general case, leading
to an {\em additional\/} constraint on the admissibility of group
actions. This will be taken up in the following two subsections, after
illustrating the above considerations by means of the elementary
systems $T^*S^1$ and $T^* \dR^+$.  Thereafter we will add some remarks
on the quantization of general cotangent bundles $T^*Q$ and subbundles
including our example system $\CS$.

\subsection{Application to $T^* \dR^+$ and $T^*S^1$}

To illustrate the quantization scheme reviewed in the previous
subsection, we present the examples $T^* \dR^+$ and $T^*S^1$. The
second of these examples will lead us to discuss the issue of
generating phase space functions in more detail. 

As $T^*S^1$ and $T^* \dR^+$ may be quantized by various established
quantization schemes, we present the standard results (from several
perspectives) first before turning to their group theoretical
quantization.

\subsubsection{Standard results}
\label{Standard}

As remarked already in the preceding subsection, there is a
one--parameter family of different, but physically acceptable quantum
theories of $T^* S^1$. This parameter may be viewed as a consequence
of the multiple connectedness of the phase space (cf, e.g., the
general statements on the quantization of multiply connected phase
spaces in geometric quantization in Ref.~\cite{Woodhouse} and our
discussion below; cf also Ref.~\cite{Nico}).  The resulting
Hilbert space may be spanned by the wave functions $\exp
i(n+\theta)\varphi$, $n \in \dZ$, where $\varphi$ is a coordinate on
the interval $[0,2 \pi]$ and $\theta \in [0,1]$ (where $\theta =0$ is
to be identified with $\theta=1$) is the fixed parameter mentioned
above. Thus the wave functions may be regarded as functions on the
interval $[0,2 \pi]$ with quasi--periodic boundary conditions (having
periodic probability densities). The momentum operator $p=(\hbar/i) \,
d/d\varphi$ is self--adjoint and its spectrum obviously is of the form
$\{ \hbar(n + \theta), n \in \dZ\}$.  (The level spacing $\hbar$ of
the spectrum is fixed by the choice $[0,2\pi]$ for the fundamental
interval of the angle variable $\varphi$.  In physical applications,
it may possibly be rescaled depending on the realization of
$\varphi$.)  The spectra of $p$ differ for different values of
$\theta$ and thus the respective quantum theories cannot be unitarily
equivalent. (Note also that although $\theta$ provides only an overall
shift in the spectrum of $p$, already for a free Hamiltonian of the
form $H=p^2/2$, energy differences {\em are}\/ affected by that
parameter.)

From the point of view of geometric quantization (cf, e.g.,
\cite{Woodhouse}) wave functions are sections in a line bundle over
$S^1$ or better $T^*S^1 = S^1 \times \dR$. This line bundle is
necessarily trivial. There are, however, several inequivalent
connections $\hbar^{-1}\Theta$ with the same curvature $\hbar^{-1}
\omega$ ($\omega$ being the symplectic form on $T^*S^1$; its Chern
class is trivial and so is the bundle). Up to gauge transformations
$\Theta$ may be brought into the form $\Theta = p d\varphi -
\theta\hbar d\varphi$ where $\theta \sim \theta + 1$ as a consequence
of the $U(1)$ gauge transformations $\exp(-i \varphi)$. The difference
between two connections with fixed curvature is a flat connection; up
to gauge transformations this difference is an element of
$H^1(M,\dR)/H^1(M,\dZ)\sim H^1(M,U(1))$ ($M$ being the phase space
under consideration, here $M= T^*S^1$), different choices correspond
to different parallel transporters around nontrivial loops (so the
ambiguity may be associated also to elements of
${\mathrm{Hom}}(\pi_1(M),U(1)$).  In the above trivialization of the
bundle, (polarized) wave functions correspond to ordinary functions
over $S^1$ and the $\theta$--angle enters in the momentum operator
$\hat p$: $\hat p = -i \hbar \nabla_{X_p} + p = -i \hbar
\frac{d}{d\varphi} + \theta\hbar$ (here $X_p=\frac{d}{d\varphi}$ is the
Hamiltonian vector field corresponding to $p$ and $\nabla$ denotes the
covariant derivative).  We may, however, also use a nontrivial
transition function at $\varphi = 0$ to remove $\theta$ in the
expression for the momentum operator, transferring it simultaneoulsy
into the wave functions (so that effectively they are quasiperiodic as
above).

The quantum system corresponding to $T^* \dR^+$, on the other hand,
may be traced back to the one for $T^*\dR$: Let $q >0$ and $p \in \dR$
parameterize $(T^* \dR^+, dq \wedge dp)$. Then in the new chart
$(\widetilde q := \ln q$, $\widetilde p := q \, p)$ this symplectic
manifold {\em becomes}\/ just $(T^* \dR, d\widetilde q \wedge
d\tilde p)$. The quantization of the latter is standard, yielding
wave functions $\widetilde{\psi}(\widetilde q)$ with measure $\int \!
d \widetilde q$ if simultaneously $\widetilde p = (\hbar/i) \,
d/d\widetilde q$.

The resulting quantum system may now be represented also in the
original coordinates $q = \exp (\widetilde q)$: For the wave functions
$\psi(q) = \widetilde{\psi}( \ln q )$ the measure in the inner product
becomes $\int_{\dR^+} dq/q$ and $\widetilde p = (\hbar/i) \,q \, d/d
q$. Note that, as a consequence of the nontrivial measure in $q$, $p =
(\hbar/i) \, \sqrt{q} \, (d/d q) (1/\sqrt{q})$.

We can also get rid of the nontrivial measure by rescaling the
wavefunctions: $\widehat \psi (q) := \psi (q)/\sqrt{q}$.  Then the
measure becomes $\int_{\dR^+} dq$, while $p= (\hbar/i) \, d/d q$ and
now $\widetilde p$ is seen to become $\widetilde p = (qp + pq)/2$.

We remark that while $\widetilde p$ is (in the above manner by
construction) a self--adjoint operator, $p$ is only hermitean, having
no self--adjoint extensions (cf Ref.~\cite{Thirr3Aufl2} and
Ref.~\cite{Isham} for further details on $p$).

\subsubsection{$T^*\dR^+$}
\label{s:TRp}

We next turn to the group theoretical quantization of $T^*\dR^+$. As
remarked already above, from the point of view of symplectic manifolds
$T^* \dR^+=T^*\dR$. The {\em observables}\/ $q>0$ and $p$ on $T^*
\dR^+$, however, are certainly different from the observables $q$ and
$p$ on $T^*\dR$, which, for means of clarity, we again denote by
$\widetilde q$ and $\widetilde p$ as in Subsec.\ \ref{Standard} above.
The precise correspondence between these observables (viewed as
observables on {\em one and the same\/} phase space) has been provided
already there, too.

In the group theoretical approach there are thus at least two
admissible groups which may be used to quantize $T^*\dR^+$ (or,
likewise, to quantize $T^*\dR$). First, we may just take the abelian
group generated by the Hamiltonian vector fields corresponding to
$\widetilde q \equiv \ln q$ and $\widetilde p \equiv qp$. In
one--to--one correspondence with the quantization of $T^*\dR$, this
action on $T^*\dR^+$ has no momentum map, and the canonical group $\CC$
becomes the three--dimensional Heisenberg group. As is evident from
the discussion of $T^*\dR^+$ in the preceding Subsec.\ \ref{Standard}, in
this way the correct quantum theory of $T^*\dR^+$ is reproduced. It is
identical to the quantum theory of $T^* \dR$; one just has to take
into account the nontrivial correspondence of observables.

Second, in the framework of group theoretical quantization, we may
also use the group generated by $q$ and $\widetilde p \equiv qp$. This
is easily seen to provide an effective and transitive action on
$T^*\dR^+$ of the two--dimensional, nonabelian affine group $\CG = \dR
\semidir \dR^+$.  Since it obviously has a momentum map, for the
quantization of $T^*\dR^+$ (or, likewise, also of $T^*\dR$!) one may
study, as an alternative to the Heisenberg group $\CC$, the unitary
IRREPs of the affine group $\CG$.

There are {\em three}\/ unitarily inequivalent IRREPs of $\CG$ (again
here we refer to Ref.~\cite{Isham} for further details). In one of them,
the operator $q$ has a strictly negative spectrum; clearly, this
representation has to be excluded on physical grounds, as classically
$q$ is strictly positive. Furthermore, one of the representations uses
a {\em one}\/--dimensional Hilbert space and thus does not come into
question as quantum theory of $T^*\dR^+$, too.  The single remaining
representation has the Hilbert space $L^2(\dR^+,dq/q)$ with the
generator $\widetilde p$ being represented by $(\hbar/i) \, q \,
(d/dq)$. This is in coincidence with what we found above (cf Subsec.\
\ref{Standard}). In this case the parameter $\hbar$ enters on reasons of
correct physical dimensions: $\widetilde p$ has the dimension of an
action and the Poisson bracket relation $\{q,\widetilde p\}=q$ thus
has to turn into the commutator $[q,\widetilde p]=i\hbar q$ in the
quantum theory.

Certainly $\{\cdot,q \}=-d/dp$ and $\{\cdot,p\}=d/dq$ do {\em not}\/
generate a group on $T^*\dR^+$; $d/dq$ generates translations of $q$,
which may leave the positive real axis. This fits well to the
previous observation that $p$ cannot become a self--adjoint operator.

\subsubsection{$T^*S^1$}
\label{TS1}

In the group theoretical approach to quantizing $T^*S^1$ one first
looks for a transitive, almost effective Hamiltonian action of a group
$\CG$ on that space. Such a group is provided by the
three--dimensional Euclidean group (in two dimensions) $E_2=\dR^2
\semidir SO(2)$. If $\varphi \in [0,2\pi]$ denotes the configuration
space variable on $S^1$ and $p$ its conjugate momentum, Hamiltonian
generators of this action are provided by $\{\cdot,p\}$, generating
rotations along the $S^1$, as well as by $\{\cdot, \sin \varphi\}$ and
$\{\cdot, \cos \varphi\}$, which generate transformations along the
fibers $\varphi =$ const. Since the Poisson brackets between the
respective Hamiltonian functions clearly close, the action has a
momentum map. The action of $E_2$ is easily seen to be effective and
transitive on $T^*S^1$, moreover. The representation theory of $E_2$
shows that there is a one--parameter family of unitary IRREPs (cf
Ref.~\cite{Isham} for details). The corresponding parameter $\lambda
\in \dR^+$ is, however, {\em not}\/ the $\theta$--angle, as we might
have expected from our previous consideration of this example.
Instead, upon working with dimensionful quantities, it may be seen
that this parameter has to be identified with $\hbar$ again.

In the present quantization scheme the $\theta$--angle arises only
when considering {\em another}\/ group action on $T^*S^1$. Clearly
$\pi_1(E_2)=\pi_1(SO(2))=\dZ$. Thus instead of $E_2$ we may consider
as well the action of its universal covering group, $\widetilde E_2$.
This action is no more effective, but still almost
effective\footnote{The elements which act trivially on $T^*S^1$ are
  then just the center $\dZ$ of $\widetilde E_2$  in the kernel of
  the   projection from $\widetilde E_2$ to $E_2$.} (and the Lie algebra
isomorphism between the Poisson algebra of the generating Hamiltonians
and the elements of $\CL(\CG)$ is certainly not affected by this
change of $\CG$). The unravelling of the subgroup $SO(2)$ of $E_2$ to
$\dR \subset \widetilde E_2$ leads to an additional continuous
parameter in the unitary representations. This parameter lives on a
circle and may be identified readily with the angle $\theta$. So, when
using $\CG=\widetilde E_2$, the group theoretical quantization scheme
reproduces the general results of other established approaches. (The
representations obtained from the choice $\CG = E_2$ correspond to the
special, but still legitimate, quantum realization with periodic wave
functions, $\theta =0$, on the other hand).

\subsubsection{Summary}

The two examples discussed above nicely illustrate that there may be
several admissible group actions on one and the same phase space
$\CP$ (which will still be the case also after imposing our
additional condition on admissible group actions below).

As any covering group $\widehat G$ of an (almost) effectively acting
group $G$ acts almost effectively on the phase space, too, and unitary
representations of $G$ are also unitary representations of $\widehat
G$ (but not necessarily vice versa), we will always choose the
(unique) simply connected universal covering group $\widetilde G$ as
the group $\CG$.

We learn from the group theoretical quantization of $T^*S^1$ that only
then we may expect to obtain the most general quantum realization of
the theory with classical phase space $\CP$.

The example $T^* \dR^+$ (or $T^*\dR$) demonstrates that different
admissible groups need not be just coverings of one another.
Moreover, this example illustrates that not all weakly continuous,
unitary IRREPs of $\CG$ need to make sense physically. In part this
was concluded from a comparison of the range of values of a physically
important classical observable (namely $q$) with its quantum spectrum.

The situation in quantizing the phase space $\CS=S^1\times \dR^+$,
discussed in detail in Sec.\ \ref{s:S}, will be quite analogous to the
one in quantizing $T^*S^1$.  The allowed effectively acting group will
be $SO^{\uparrow}(1,2)$. Only by studying the IRREPs of the respective
universal covering group a $\theta$--angle, to be expected due to
$\pi_1(\CS)=\dZ$, will be obtained. In analogy to the example
$T^*\dR^+$, on the other hand, not all unitary IRREPs will be seen to
make sense ``physically'' as quantum realizations of $\CS$.

\subsubsection{Other group actions on $T^*S^1$}

Up to now the discussion was in agreement with Ref.~\cite{Isham}.
However, in the example of $T^*S^1$ there are much more group actions
which fulfill the conditions of transitivity, effectiveness, and of
being Hamiltonian with momentum map: The Lie algebra of $E_2$ is not
only provided by the Hamiltonian generators $\{\cdot,p\}$,
$\{\cdot,\sin\varphi\}$, and $\{\cdot , \cos\varphi\}$ on $T^*S^1$,
but also by the countably infinite family $\{\cdot, l^{-1}p\}$,
$\{\cdot, \sin l\varphi\}$, and $\{\cdot, \cos l\varphi\}$,
$l\in\dN$.\footnote{This example, for which we are grateful to
  H.~Kastrup, provides a simplified version of what will be found for
  the phase space $\CS$ by the systematic procedure employed in Sec.\ 
  \ref{s:S} (cf Eqs.\ (\ref{moment}) and (\ref{Lieiso1}) below).} For
fixed $l$ these vector fields generate an {\em effective}\/ action of
the $l$--{\em fold covering}\/ group of $E_2$: The vector field
$l^{-1}\{\cdot,p\}$ generates the translations
$\varphi\mapsto\varphi+tl^{-1}$, $t \in \dR$, which is the identity
transformation for $t=2\pi l$, but not already for $t=2\pi j$, $j<l$.

If we repeat the quantization described in Subsection~\ref{TS1} for
$l\not=1$, we have to use the same representation theory because in
any case we use the universal covering group $\widetilde{E}_2$.
However, now we have $l^{-1}p$ in place of $p$, and {\em this} phase
space function is quantized to the same operator as $p$ above with
discrete spectrum $\hbar(\dZ+\theta)$.  Thus $p$ will be quantized to
an operator with spectrum $\hbar l(\dZ+\theta)$.  Note that the
interval $\varphi\in[0,2\pi]$ has not changed and, therefore, the
obtained spectrum is not acceptable. A rescaling of $\hbar$ to absorb
$l$, furthermore, is not possible because locally we have to preserve
canonical conjugacy of $p$ and $\varphi$.

We are thus in the need of excluding the group actions on $T^*S^1$ with
$l \not = 1$! To extract a general strategy from this example, we now
will focus on the question of what kind of phase space functions
may be generated by the fundamental observables of the group action.

\subsection{Generation by fundamental observables}
\label{Generating}
In Ref.~\cite{Isham} two different principles for what
phase space functions can be generated by the fundamental observables
$F_1,\ldots,F_n\in C^{\infty}({\cal P},\dR)$ generating the group
action on the phase space were presented:

\medskip

\noindent {\bf Strong Generating Principle (SGP):} {\it For any phase space
  function $f\in C^{\infty}({\cal P},\dR)$ there is a function
  $\Phi_f\in C^{\infty}(\dR^n,\dR)$ such that
  $f=\Phi_f(F_1,\ldots,F_n)$.}

\medskip

\noindent {\bf Local Generating Principle (LGP):} {\it Any $s\in{\cal P}$
  has a neighborhood ${\cal U}_s\subset{\cal P}$ such that the condition of
  the SGP is met on ${\cal U}_s$.}

\medskip

As noted in Ref.~\cite{Isham}, the LGP is fulfilled if the group action is
transitive. However, transitivity is not sufficient for SGP. E.g., in
the example of $T^*S^1$ above we found an infinite family of
transitive group actions parameterized by the label $l$. The SGP is
fulfilled only for $l=1$: For $l>1$ the functions $\sin l\varphi$ and
$\cos l\varphi$ are not sufficient to generate an arbitrary (smooth)
function on the interval $0\leq\varphi<2\pi$, because any generated
function is $2\pi l^{-1}$--periodic (globally we cannot take the
$l$--th root).  Thus, demanding SGP singles out the only group action
which reproduces the results of standard quantizations in this
example.

In Ref.~\cite{Isham} only the need for the LGP was recognized, and
incorporated by means of transitivity of the group action.
(Consequently it then was concluded \cite{Isham}, p.\ 1149: ``... in
this group theory oriented quantization scheme, we cannot always
maintain the strong generating principle.'') As the above example
shows, however, the validity of the SGP is an essential part of group
theoretical quantization and must not be ignored.

The SGP is a condition on the Hamiltonians of a given group action.
For practical applications it may be worthwhile to reformulate it in
terms of a property of the group action (analogously to trading in
transitivity of the action for the LGP) or even the canonical group
$\CG$ itself. We did not succeed in this attempt in full generality.
However, we will now present a {\em necessary\/} condition for the
validity of the SGP for a rather large class of group actions.

Let $\CG$ have an almost effective, transitive Hamiltonian action on
$\CP$ and let us, for the above purpose, assume that this action
admits a momentum map (and thus there is no need for a central
extension).  An almost effective action can always be reduced to an
effective action by factoring out a discrete subgroup: If $\CG$ acts
almost effectively, then $G:=\CG/N$, where $N$
is the maximal invariant subgroup of $\CG$ acting
trivially on $\CP$, acts effectively (with all other properties of the
action unchanged). The necessary condition mentioned above may now be
formulated as a condition on the remaining center $Z(G)$ of $G$.

\begin{lemma}\label{generating}
  Let $\CG$ be a group acting almost effectively, transitively, and
  Hamiltonian with a momentum map on the phase space ${\cal P}$ and $G$
  be the corresponding effectively acting group. If $\CG$ is
  semisimple and the center $Z(G)$ of $G$ is nontrivial, then the
  strong generating principle is violated. It is also violated (for a
  general group $\CG$), if $Z(G)$ is nontrivial but finite.
\end{lemma}

\begin{proof}
  Let $s\in{\cal P}$, $g\in G$, and denote the group action of $g$ by
  $L_g\colon s\mapsto gs$. By means of this action to each $X\in\CL G$
  a vector field $\widetilde{X}$ on ${\cal P}$ is associated, whose
  flow we denote as $\exp tX:=\Phi_t(\widetilde{X})$.  Its pushforward
  with $L_g$ acting on a function $f$ on ${\cal P}$ is
\begin{equation}
  L_{g*}\widetilde{X}(f) = \left.\frac{d}{d t}\right|_{t=0}f\circ
  L_g\circ\exp(tX)=
   \left.\frac{d}{d t}\right|_{t=0}f\circ\exp(tAd_gX)\circ L_g\, .
\end{equation}
If $\widetilde{X}=\widetilde{X}_H=\{\cdot,H\}$ is a Hamiltonian vector
field, then we have, furthermore,
$L_{g*}\widetilde{X}_H=\widetilde{X}_{H\circ L_g^{-1}}$ because the
group action is Hamiltonian. For $g=z\in Z(G)$ in the center of $G$ we
have $Ad_gX=X$ and these equations imply $\widetilde{X}_H=
\widetilde{X}_{H\circ L_z^{-1}}$.  The generating function $H$ thus
has to fulfill
\begin{equation}\label{ztrans}
  H\circ L_z^{-1}=H+c_z(H)\quad\mbox{ for any }z\in Z(G).
\end{equation}
Here $c_z\colon{\cal L}G\to\dR$ is a linear map from the Lie algebra
of $G$, which we identify using the momentum map with its isomorphic
Lie algebra of generating functions of the group action on ${\cal P}$,
to $\dR$.  This map is in fact a 1--cocycle in the cohomology of this
Lie algebra: $\{H\circ L_z^{-1},G\circ L_z^{-1}\}=\{H,G\}\circ
L_z^{-1}$ implies $\{H+c_z(H),G+c_z(G)\}=\{H,G\}=\{H,G\}+c_z(\{H,G\})$
which leads to $c_z(\{H,G\})=0$. This observation already proves our
first assertion: If $G$ is semisimple, we have $[{\cal L}G,{\cal
  L}G]={\cal L}G$ and $c_z({\cal L}G)=c_z([{\cal L}G,{\cal L}G])=0$;
$c_z$ vanishes for any $z\in Z(G)$. This means that each of the
generating functions, and therefore any generated function, is
invariant with respect to the action of $Z(G)$. But the center of $G$
acts nontrivially, because the group action of $G$ is effective, and
not any phase space function, which is in general not invariant, can
be generated.

For groups with $[{\cal L}G,{\cal L}G]\not={\cal L}G$ (nonperfect
groups, cf the remark following this proof) the above argument
cannot be used.  However, if $Z(G)$ is finite, there is for each $z\in
Z(G)$ a $k\in\dN$ with $z^k=1$. Due to $H=H\circ
L_{z^k}^{-1}=H+kc_z(H)$ (This follows from Eq.~(\ref{ztrans}) and
$c_z(H\circ L_{z'}^{-1})=c_z(H)$ for all $z,z'\in Z(G)$, which in turn
is a consequence of $c_z(H+c)=c_z(H)$ for any constant function $c$ on
the phase space.) we again have $c_z(H)=0$ for any $z\in Z(G)$ and
$H\in{\cal L}G$.
\end{proof}

In the above lemma, we could also relax the conditions replacing
``semisimple'' by ``perfect''.\footnote{We are grateful to D.\ Giulini
  for this remark.}  The defining property of a perfect
Lie group $\CG$ is $[{\cal L}\CG,{\cal L}\CG]={\cal L}\CG$. A
prominent example for a nonsemisimple but perfect Lie group is the
Poincar\'e group.

Note that the center of semisimple Lie groups is discrete, while for
perfect Lie groups per se this is not necessarily the case. However,
in the present context a continuous center $Z(\CG)$ is excluded in any
case due to the (almost) effectiveness of the $\CG$--action and the
existence of a momentum map: The phase space function generating the
action of the center would have vanishing Poisson brackets with all
other generating functions.  Therefore, it would be constant, and the
center would act trivially.

Thus, the only case of a nontrivial center not covered by the lemma is
that of a discrete but infinite center of a nonperfect group.

A simple example for this case where, however, the SGP is still
violated, may be provided on $T^*\dR$. Such an action on $T^*\dR$
fulfilling Isham's axioms can be constructed as a limit $l\to\infty$
of the action of the $l$--fold covering group of $E_2$ on $T^*S^1$:
After the symplectic transformation
$(\varphi,p)\mapsto(l\varphi,l^{-1}p)$ we can take the limit
$l\to\infty$ for the action of the $l$--fold covering group of $E_2$.
The generating functions $p$, $\sin\varphi$, and $\cos\varphi$ are now
$l$--independent, but the $l$--fold covering group acts on a phase
space with $\varphi$--interval $0\leq\varphi<2\pi l$. For $l\to\infty$
this phase space unwinds to $T^*\dR$ and the action becomes an
effective and transitive action of $\widetilde{E}_2$, which is neither
perfect nor has finite center.  The lemma does not apply, but
nevertheless the group action has to be rejected because only
$2\pi$--periodic functions can be generated.

This example shows that the lemma is not sufficient to decide in all
cases whether a group action is allowed, and it demonstrates even more
drastically the necessity of the SGP: Trusting this group action of
$\widetilde{E}_2$ would lead us to a discrete spectrum for $p$ in a
quantization of $T^*\dR$! (This discreteness comes in because the
fundamental observables are periodic, which is a global property and
cannot be detected by the LGP. A further failure of this group action
is that the coordinate $q$ in $T^*\dR$ could not be promoted to an
operator, because it cannot be generated by the fundamental
observables.)

Note that the lemma does not provide any statement about the validity
or failure of the SGP for the case that $Z(G)$ is trivial. We are,
however, not aware of an example with trivial $Z(G)$ where the SGP is
violated.

\vspace{1.5mm}

In the paragraph preceding the lemma we made use of the fact that a
trivially acting subgroup of $\CG$ can always be factored out to
arrive at the effectively acting group $G$. If the center of the latter
group, $Z(G)$, is nontrivial, it can be factored out only at the cost
of factoring the phase space, too.  This does not change its
dimensionality due to discreteness of the center. (To do so, we have
to suppose that the action on the phase space of the center is
properly discontinuous, which is, e.g., fulfilled if the center is
finite.) If the action of $\CG$ on this factored phase space is still
Hamiltonian, the conclusion of the lemma can be evaded by regarding
$\CG$ as canonical group for this smaller phase space $\CP'
\equiv \CP/Z(G)$ (i.e.\ although the SGP is violated on $\CP$ it is
not necessarily so on $\CP'$). 

In the light of this consideration we can understand the wrong
$p$--spectrum obtained when using the action of the $l$--fold ($l>1$)
covering group of $E_2$ on $T^*S^1$. In this case the center is the
cyclic group of order $l$ generated by the translation in $\varphi\in
[0,2\pi]$ by $2\pi l^{-1}$.  If we want to factor out the center, we
have to identify the points $\varphi$ and $\varphi+2\pi l^{-1}$ to
obtain an action of $E_2$ (or an almost effective action of the
$l$--fold covering).  This identification effects a reduction of the
configuration space to the interval $[0,2\pi l^{-1}]$, which explains
the multiplication of the $p$--spectrum by $l$.

\subsection{Quantizing cotangent bundles and certain subbundles}
\label{s:sub}

We proceed with some general remarks \cite{Isham} on the
group theoretical approach when applied to phase spaces which are
cotangent bundles, $\CP = T^* Q$.  As discussed in the next section,
the phase space $\CS$, on the other hand, is definitely not a
cotangent bundle.  However, it will turn out to be a certain subbundle
of a cotangent bundle (specified below).  Many of the facts applicable
to cotangent bundles will be seen to be applicable to those
subbundles, too. In a sense, the situation even simplifies there.

\subsubsection{A general strategy for determining group actions}
\label{s:general}

On $T^*Q$, the infinite dimensional group $\CD :=
\left(C^\infty(Q,\dR)/\dR\right) \semidir \Diff(Q)$, which is a
subgroup of the full group of canonical transformations, always acts
transitively and effectively.

Here $\Diff(Q)$ is the canonical lift of the diffeomorphism group of
the configuration space $Q$.  If $q^i \to \widetilde{q}^i(q)$ denotes
the diffeomorphism on $Q$, this is lifted canonically to a
symplectomorphism on $T^* Q$ (a so--called ``point transformation'')
when it is accompanied by $p_i \to p_j \, \partial q^j/\partial
\widetilde q^i$ (where $q(\widetilde q)$ denotes the inverse of the
function $\widetilde q(q)$). The action of $\Diff(Q)$ is also
Hamiltonian and allows a momentum map: If $X^i(q) \, d/dq^i$ is the
generating vector field of a diffeomorphism of $Q$ (connected to the
identity), then $X^i(q) \, p_i$ is a Hamiltonian of its canonical
lift, and it is obvious that the Poisson algebra of these functions on
$T^*Q$ is closed without a central extension.

Although infinite--dimensional, $\Diff(Q)$ by itself does not act
transitively on $T^*Q$, as the (dim$(Q)$--dimensional) subspace
$p_i=0$ is mapped into itself.  However, when enhanced by
$C^\infty(Q,\dR)/\dR$ (``diffeomorphisms up the fibers''), the action
becomes transitive on $T^*Q$; here $C^\infty(Q,\dR)/\dR$ consists of
those canonical transformations that are generated by Hamiltonian
vector fields of the form $\{ \cdot , f(q)\}$, where $f \in
C^\infty(Q,\dR)/\dR$ ($\dR$ corresponding to the constants that act
trivially and which are thus removed so as to obtain an effective
action).

Quantizing (finite--dimensional) cotangent bundles, one thus may look
for {\em finite}\/--dimen\-sional subgroups $\CG = W \semidir G$ of
$\CD \equiv \left(C^\infty(Q,\dR)/\dR\right) \semidir \Diff(Q)$ which
still act transitively. As a subgroup of $\CD$ this action is then
guaranteed to be Hamiltonian and to act effectively. As seen above,
moreover, separately, each of the groups $\Diff(Q)$ and
$C^\infty(Q,\dR)/\dR$ allows a momentum map (and thus this follows
also for any of their subgroups $G$ and $W$, respectively). However,
the full (combined) group $\CD$, and thus also $\CG = W \semidir G$,
{\em may}\/ have an obstruction for a momentum map (cf the example $Q=
\dR^n$ reexamined below).

\subsubsection{The examples revisited}

In the examples discussed above we always used subgroups of $\CD$ (or
their covering groups). For $T^* \dR \sim T^*\dR^+$ this was $G=\dR$,
$W=\dR$ (which is more natural when viewing the phase space as
$T^*\dR$, the generating observables being $q$ and $p$ in the
corresponding chart) {\em or}\/ $G=\dR^+$, $W=\dR$ (more natural when
viewing the phase space as $T^* \dR^+$, the generating observables
being $q>0$ and $qp$ in this other chart). In the former case there is
an obstruction to a momentum map and one is lead to the
three--dimensinonal Heisenberg group (which is a subgroup of
$C^\infty(Q,\dR) \semidir \Diff(Q)$), in the latter case there was no
obstruction to a momentum map for $\CG = W \semidir G$. For $T^* S^1$,
on the other hand, $W=\dR^2$ and $G=SO(2)$ (the rotations along the
$S^1$) or, better, the universal covering group of the latter,
$G=\dR$.

\subsubsection{Subbundles}
\label{s:Subbundles}

We noted above that the subspace \be \label{Pnull} \CP_0 = \{ (p,q)
\in T^*Q \, | p_i=0 \quad\forall i = 1, \ldots, \mbox{dim}(Q) \} \ee
of $T^* Q$ is left invariant by the action of Diff$(Q)$. On the
(connected components of the) complement $\CP_*$ of $\CP_0$ in $T^*Q$
the action is, however, also transitive. More precisely, for
dim$(Q)=1\,$ $\CP_*$ has two connected components, which we will
denote by $\CP_+$ and $\CP_-$ for $p>0$ and $p<0$, respectively. (The
phase space $\CS$ will be found to be of this type with $Q=S^1$ in the
following section.) For dim$(Q)>1$, on the other hand, $\CP_*$ is
already connected and we have the following small lemma:

\begin{lemma}\label{CS0}
  For {\rm dim}$(Q)>1$ {\rm (dim $(Q)=1$)} the canonical lift of $\Diff(Q)$
  {\rm (}$\Diff_+(Q)$, the component of $\Diff(Q)$ connected to the
  identity) has a transitive and effective action on (the connected
  components of) $\CP_*=T^*Q \backslash \CP_0$ with a momentum map.
\end{lemma}

\begin{proof}  According to the invariance of $\CP_0$ with respect to
  $\Diff(Q)$, the action of Diff$(Q)$ does not lead out of the
  subbundle $\CP_*$.  For dim$(Q)=1$ each of the components of $\CP_*$
  is invariant only with respect to orientation preserving
  diffeomorphisms (and we thus restrict to $\Diff_+(Q)$ in this case).

  The momentum map of the action has been provided already above,
  furthermore, and its effectiveness on $\CP_*$ is obvious. Transitivity
  on $\CP_*$ ($\CP_\pm$ for dim$(Q)=1$) follows as $\Diff(Q)$
  ($\Diff_+(Q)$) acts fiber transitively on $\CP_* \subset T^*Q$
  (i.e., it acts transitively on the space of fibers), while on the
  fiber of $\CP_*$ ($\CP_\pm$) over the origin $q^i=0$ of some
  particular local coordinate system of $Q$ the vector fields
  $\{\cdot, q^i \, p_j \}$ act transitively.
\end{proof}

Note that when dealing with $\CP_*$ ($\CP_\pm$), it is not only not
necessary to add the above group $C^\infty(Q,\dR)/\dR$ (or {\em any}\/
of its subgroups) to obtain a transitive action, this is even not
possible: Already any one--dimensional subgroup of
$C^\infty(Q,\dR)/\dR$ moves points in a fiber of $T^*Q$ into its
origin $p_i=0$, so that no subgroup of $C^\infty(Q,\dR)/\dR$ yields a
group action on $\CP_*$.

Thus, if we are to quantize a phase space $\CP_*$ (or one of its
connected components), we may first search for finite--dimensional,
transitively acting subgroups $G$ of $\Diff(Q)$. Such an action of $G$
then automatically acts effectively {\em and}\/ now it also has a
momentum map, as this is the case for $\Diff(Q)$. The quantum
realizations of the phase space $\CP_*$ are then to be found among the
unitary IRREPs of $\CG = \widetilde G$, where $\widetilde G$ is the
universal covering group of $G$.  This sets the strategy for what
follows in the next section.

There certainly is no guarantee that such a finite--dimensional group
$G$ exists for a given (finite--dimensional) phase space $\CP_*$ as
likewise there need not exist a finite--dimensional, transitively
acting subgroup of $\CD$ on a cotangent bundle $T^*Q$. In both of
these cases there still could be some other finite--dimensional
subgroup $\CG$ of the {\em full}\/ group of canonical transformations
on the phase space acting transitively and effectively.  Moreover,
certainly not any finite--dimensional cotangent bundle (and likewise
not any of its subbundles $\CP_*$) can be quantized by the group
theoretical approach (using {\em fininte--dimensional\/} groups), even
if it is quantizable e.g.\ in the sense of geometric quantization. In
particular, the mere existence of a transitive, almost effective
action of $\CG$ on a phase space $\CP$ implies that (topologically)
$\CP \cong \CG/\CH$, where $\CH$ is a subgroup of $\CG$ (the
stabilizer group of some point in $\CP$); clearly not any phase space
$\CP$ (or also cotangent bundle $T^*Q$ or its subbundles $\CP_*$) has
the topology of some coset space of finite dimensional
groups.\footnote{We are grateful to D.\ Giulini for pointing out to us
  that {\em any\/} manifold can be obtained as the coset space of
  appropriate, generically {\em infinite dimensional\/} groups.}
Still, the group theoretical quantization scheme, and in particular
the above strategy for quantizing $T^*Q$ and $\CP_*$, is general
enough to be applicable to a number of physical systems, and, among
others, this will apply also to the phase space $\CS$.

\subsubsection{$T^*\dR^+$ and $T^*(\dR^2\backslash\{(0,0)\})$ as subbundles}
\label{s:TRpSub}

The phase space $T^*\dR^+$ can, after interchanging $q$ and $p$, be seen
as a subbundle $\CP_+$ of $T^*\dR$. A transitive action on $\dR^+$ is
generated by the phase space function $q$, whereas the proof of
Lemma~\ref{CS0} suggests to use in addition the generating function
$qp$ to obtain a transitive action on the phase space. This brings us
back to the group $\CG=\dR\semidir\dR^+$ of Subsection~\ref{s:TRp}.
The quantum theory obtained there was defined on the Hilbert space
$L^2(\dR^+,dq/q)$ with $qp$ acting as $(\hbar/i)\,q\,d/dq$, which realizes
the unique representation of $\CG$ having positive spectrum for $q$.

An example for a phase space $\CP_*$ is, again after interchanging
coordinates and momenta, the phase space
$T^*(\dR^2\backslash\{(0,0)\})$. Such a phase space is of relevance in
the context of the Aharanov Bohm effect.

The smallest transitively
acting subgroup of the diffeomorphism group of $\dR^2$ (the fibers of
this phase space) is the two--dimensional abelian group of
translations generated by the coordinates $x$ and $y$ of
$\dR^2\backslash\{(0,0)\}$. According to the proof of Lemma~\ref{CS0}
we obtain a transitive action on the phase space if we add the
functions $xp_x$, $xp_y$, $yp_x$ and $yp_y$ as generators. However,
already the span $\langle x,y,xp_x+yp_y,xp_y-yp_x\rangle$ (the latter
two functions are $xp_x+yp_y=rp_r$ and $xp_y-yp_x=p_{\varphi}$ in
polar coordinates) is closed under Poisson brackets forming a Lie
algebra isomorphic to $\dR^2\semidir\dR^2$, and we will see that it
generates a transitive action on $T^*(\dR^2\backslash\{(0,0)\})$.

The Hamiltonian vector fields are easily seen to generate an action of
the group $G=(\dR^2\semidir SO(2))\semidir\dR^+$, the semidirect
product of the group of motions of $\dR^2$ with the group $\dR^+$ of
dilatations with composition $(\vec{v}_1,R_1,\lambda_1)
(\vec{v}_2,R_2,\lambda_2)=
(\vec{v}_1+\lambda_1R_1\vec{v}_2,R_1R_2,\lambda_1\lambda_2)$.  (Here
$\vec v$ denotes the translation vector and $R$ the two--by--two
rotation matrix.)  This group is isomorphic to
$G\cong\dR^2\semidir(SO(2)\times\dR^+)\cong \dC\semidir\dC^*$ with
composition $(\alpha_1,\beta_1)(\alpha_2,\beta_2)=
(\alpha_1+\beta_1\alpha_2,\beta_1\beta_2)$. Using the latter form, the
action on $T^*(\dR^2\backslash\{(0,0)\})$ can most compactly be
written in terms of the complex coordinates $z:=x+iy$, $p:=p_x+ip_y$
as $(\alpha,\beta)\colon (z,p)\mapsto (\beta z,\beta^{-1}p+\alpha)$.
(The group $G$ can also be viewed as a subgroup of $\CD$ in Subsec.\ 
\ref{s:general}, where $\dR^2$ is a subgroup of
$C^{\infty}(Q,\dR)/\dR$ and $SO(2)\times\dR^+$ a subgroup of
$\Diff(Q)$ with $Q=\dR^2\backslash\{(0,0)\}$. From this point of view
one still would have to check the existence of a momentum map, which
is immediate from the present perspective of $G$ (cf Lemma 2).)

Analogously to the example $T^*S^1$ we can also find effective actions
of any covering group of $G$, but again they are excluded by the SGP. The
quantum theory, however, will be most generally provided by unitary
representations of the universal covering $\CG=\widetilde{G}$.

Using Mackey theory \cite{Mackey}, one finds that the inequivalent
(nontrivial) unitary representations of this (universal covering)
group may be presented on the Hilbert space $\CH=L^2(\dR^+\times
S^1,r\,drd\varphi)$ according to the unitary action
$(U(\vec{v},t,\lambda)\psi)(r,\varphi)= \lambda\exp(i\theta
t+ir(v_1\cos\varphi+ v_2\sin\varphi)) \psi(\lambda r,\varphi+t)$ of
$\CG$, where $t\in\dR$ is a parameter in $\widetilde G$ covering the
$SO(2)$--angle of $G$. Here $\theta \in (0,1]$ is the $\theta$--angle
expected due to $\pi_1(T^*(\dR^2\backslash\{(0,0)\}))=\dZ$, which, in
the group theoretical context, may be understood to arise from the
unitary representations of $\pi_1(G)=\dZ$, the center of $\widetilde
\CG$.  The spectra of the fundamental observables $\widehat{x}$,
$\widehat{y}$, $\widehat{rp_r}$, and $\widehat{p}_{\varphi}$ are
$\dR$, $\dR$, $\dR$, and $\dZ+\theta$, respectively. Note that here
$(0,0)$ is in the spectrum of ($\widehat{x}$,$\widehat{y}$), although
classically this point is removed from the configuration space.

\subsection{Projection quantization} 
\label{s:restriction}

By imposing the restriction to a subbundle at the quantum level we can
arrive at the quantum theory of $T^*\dR^+$ also in a different way:
Starting from the standard quantization of $T^*\dR$ on the Hilbert
space $\widetilde{\CH}=L^2(\dR,dq)$ we restrict it, in a second step,
to the maximal subspace $\CH$ on which $q$ is quantized to a positive
operator (implementation of the restriction $q>0$ at the quantum level
as an operator inequality), i.e., we define $\CH$ through completion
of the maximal subspace $F\subset\CD(\hat{q})$ on which
$\int_{\dR}\overline{f}qf\,dq\geq 0$ for all $f\in F$ (where
$\CD(\hat{q})$ is the domain of definition of the multiplication
operator $\hat{q}$). This subspace is easily seen to be
$\CH=L^2(\dR^+,dq)$.

Clearly there is a (unique) projector $\pi\colon \widetilde \CH \to
\CH$, which may be used to also transport operators defined in
$\widetilde \CH$ to operators on $\CH$. (The uniqueness of the
projector is a result of the maximality condition required for the
subspace $\CH$ on which $\widehat q>0$. This condition is necessary to
reproduce standard results; in a way, it serves to capture 
the phase space, here $T^*\dR^+$, globally.)

We now propose a more general setting in which the above ``projection
quantization'' 
should be applicable.

\subsubsection{Restricted phase spaces and their Hilbert spaces}

A phase space $\CP$ which can be treated using projection quantization 
has to obey the following properties: First, $\CP$ can be
characterized as a submanifold of a phase space $\widetilde \CP$ via
restriction by means of inequalities $f_i>0$ for a set of functions
$\{f_i\}$ on $\widetilde \CP$ with mutually vanishing Poisson
brackets. We furthermore demand that, for each $i$, the set on which
the opposite inequality, $f_i<0$, is fulfilled is nonempty. (This
condition is necessary to exclude, e.g., cases like the restriction of
$T^*\dR^2$ to $T^*(\dR^2\backslash\{(0,0)\})$ by means of $x^2+y^2>0$,
which cannot be treated by the method of the present subsection; see
the remarks below.)  Second, a quantum realization of $\widetilde \CP$
is known in which the functions $\{f_i\}$ may be promoted to
self--adjoint, simultaneously diagonalizable operators
$\{\hat{f}_i\}$.\footnote{We remark that in the case of unbounded
self--adjoint operators commutativity on a dense domain is not
sufficient for their simultaneous diagonalizability, needed below.}

For simplicity we assume that the space $\CP$, where all conditions
$f_i>0$ are fulfilled, is connected. Otherwise, we have to quantize
each connected component separately and to take eventually the direct
sum of the resulting Hilbert spaces as common Hilbert space for the
quantization of $\CP$.

The general strategy of the projection quantization 
to obtain a quantum realization of $\CP$ is then as follows: Starting
with the Hilbert space $\widetilde{\CH}$ which quantizes
$\widetilde{\CP}$ we have the self--adjoint operators $\hat{f}_i$.
Their spectral families can be used to define the projectors
$P_i:=\Theta(\hat{f}_i)$, where $\Theta:\dR\to\dR$ is the step
function which is zero for $x<0$ and one for $x\geq 0$.  Because the
operators $\hat{f}_i$ are assumed to be simultaneously diagonalizable,
their spectral families commute and the common projector $P:=\prod_i
P_i\colon\widetilde{\CH}\to\widetilde{\CH}$ can be defined
unambiguously. Using this projector, the restricted Hilbert space
$\CH$ is defined as a subspace of $\widetilde{\CH}$ according to
$\CH:=P(\widetilde{\CH})$. As a Hilbert space of its own, $\CH$ is
regarded as the Hilbert space of $\CP$.

Restricting the image of $P$ to $\CH$ we obtain a map
$\pi\colon\widetilde{\CH}\to\CH$ with adjoint being the inclusion
$\iota\colon\CH\hookrightarrow\widetilde{\CH}$ of $\CH$ (which is
defined as a subspace of $\widetilde{\CH}$) in $\widetilde{\CH}$. Both
these maps are partial isometries (i.e., they map closed subspaces ---
$\CH$ in both cases --- isometrically to their images and annihilate
their orthogonal complements). Composing the two maps we obtain
$\pi\circ\iota=\dId_{\CH}$ (the identity on $\CH$) and
$\iota\circ\pi=P$ (the projector on $\widetilde{\CH}$), respectively.

In the preceding subsection we presented a general strategy for
finding a group action on certain subbundles of cotangent bundles
appropriate for the group theoretical quantization. For a
one--dimensional configuration space the subbundle is defined by an
inequality of the form $f>0$ where $f$ is the coordinate or its
canonical momentum. All the above conditions of projection quantization 
are fulfilled in this case and it can be used to obtain a quantum
realization of this subbundle as demonstrated by the example
$T^*\dR^+$ above. 

For phase spaces of dimension greater than two the situation is
different. Here, the subbundles $\CP_*$ of the previous subsection are
defined by inequalities $f_i\not=0$ removing a {\em
  lower}--dimensional submanifold from the phase space and the
functions $f_i$ do not meet all the conditions required above.  They
still Poisson commute and for a known quantum realization of $\CP$
they correspond to simultaneously diagonalizable operators.  Thus, one
still can construct the projector, of course. However, the restriction
method may fail: If zero is not contained in the discrete part of the
spectrum of all the $\hat{f}_i$, then the projector is the identity on
$\widetilde{\CH}$, not leading to any restriction (an example for this
case is the phase space $T^*(\dR^2\backslash\{(0,0)\})$). If, on the
other hand, zero is contained in the discrete part of the spectrum for
at least one of the $\hat{f}_i$, then the projection leads to a
restriction, but the point zero can be excluded from the spectra of
all the $\hat{f}_i$ only if it is an isolated point.

Although the projection quantization 
is not applicable to
higher--dimensional phase spaces of the form $\CP_*$ in general, the
conditions for its applicability as formulated above are fulfilled by
a much wider class of systems than those considered in Subsec.\ 
\ref{s:Subbundles}. Given a phase space $\CP$ one merely has to find
an appropriate embedding of $\CP$ within a phase space with known
quantum realization. 

\subsubsection{Observables}

To complete the quantum theory of $\CP$ we have to promote a certain
class of observables to densely defined operators on $\CH$.  In the
quantization of $\widetilde{\CP}$ we already have such operators
$\widetilde{\CO}$ acting on $\widetilde{\CH}$ as quantizations of
observables. These can be used to define operators on $\CH$ by mapping
$\widetilde{\CO}\colon\widetilde{\CH}\to\widetilde{\CH}$ to
$\CO\colon\CH\to\CH$ by means of
$\CO:=\pi\circ\widetilde{\CO}\circ\iota$. If $\widetilde{\CO}$ is
densily defined with domain $\CD(\widetilde{\CO})$, then $\CO$ is also
densily defined with domain $\CD(\CO)=\pi(\CD(\widetilde{\CO}))$.

Specific properties of $\widetilde{\CO}$ are, however, not necessarily
inherited by $\CO$. E.g., an unbounded, 
self--adjoint operator $\widetilde{\CO}$ leads, in general, only to a
hermitean operator $\CO$: The product of adjoints of two densely
defined operators $A\colon F\to G$ and $B\colon G\to H$ between
Hilbert spaces satisfies $A^*B^*\subset(BA)^*$, and equality can be
concluded, without further information on $A$ and $B$, only if $B$ is
bounded (defined on all of the Hilbert space $G$ and not just on a
dense subset). This condition is fulfilled for the maps $\pi$ and
$\iota$ in the definition of $\CO$, such that we obtain as its adjoint
$$
  \CO^*=(\pi(\widetilde{\CO}\iota))^*=(\widetilde{\CO}\iota)^*\pi^*\supset
  \iota^*\widetilde{\CO}^*\pi^*= \pi\widetilde{\CO}^*\iota \; .
$$
If $\widetilde{\CO}$ is self--adjoint,
$\widetilde{\CO}=\widetilde{\CO}^*$, then $\CO$ is in general only
hermitean: $\CO\subset\CO^*$. (Cf also the example of the momentum
operator of $T^*\dR^+$ below.)

Similarly, for a unitary operator $\widetilde{\CO}$ the operator $\CO$
is isometric ($\pi \widetilde{\CO} \iota$ clearly preserves the norm
on $\CH$), but not necessarily also unitary: Its adjoint is given by
$\CO^*=\pi\widetilde{\CO}^*\iota$ due to the fact that
$\widetilde{\CO}$, being unitary, is a bounded operator. Only if
$\widetilde{\CO}$ commutes with $P$ (i.e.\ if $\widetilde{\CO}$
preserves the subspace $\CH=P(\widetilde{\CH})$ as well as its
orthogonal complement), we may in general simplify
$\CO\CO^*=\pi\widetilde{\CO}P\widetilde{\CO}^*\iota$ to $\CO\CO^*=\pi
P\iota=\dId_{\CH}$ and likewise conclude $\CO^*\CO=\dId_{\CH}$.  An
example for a unitary operator with only isometric projection will
appear in Subsec.\ \ref{s:Restrict}.

If possible, observables of the classical theory are promoted to
self--adjoint operators. (The momentum operator on $T^*\dR^+$ provides
an example where this is not possible.)  The operator $\CO$ obtained
from some self--adjoint operator $\widetilde{\CO}$ in the above manner
is, in general, only hermitean; this is typically the case because the
conditions $f_i>0$ introduce a boundary on the phase space $\CP$. An
operator $\CO$ projected as above is then defined on a dense domain
including a specification of boundary conditions. If this operator has
self--adjoint extensions, each of them can be used as quantization of
an observable (possibly introducing an additional ambiguity in
defining the quantum theory of $\CP$).

The latter scenario may be illustrated by means of a particle on a
line of bounded extension. This system may be obtained as a
submanifold of $T^*\dR$ by means of $f_1 = q -a$ and $f_2=b-q$ for
some $a,b \in \dR$ with $b>a$. The domain of definition of the
momentum operator projected from the one of $T^*\dR$ is given by
absolutely continuous functions on $[a,b]$ which vanish at the
boundary. So defined, it is only hermitean. However, it has a family
of self--adjoint extensions, parameterized again by a $\theta$--angle,
defined on absolutely continuous functions $\psi$ satisfying
$\psi(b)=\exp(i\theta) \, \psi(a)$. Each of these extensions may now
be chosen as a possible quantum observable corresponding to the
canonical momentum on $T^*([a,b])$ (cf, e.g., Ref.~\cite{Thirr3Aufl2}). 

Best candidates for operators which project to a self--adjoint one on
$\CH$ correspond to phase space functions adapted to the boundary.
This is similar to the situation in group theoretical quantization,
where the condition that the fundamental observables generate an
action on $\CP$ forces the generating vector fields to be tangential
to the boundary.

We finally illustrate these considerations by means of the
quantization of $T^*\dR^+$. The Hilbert space $\CH=L^2(\dR^+,dq)$ was
derived at the beginning of this subsection 
using the projection quantization. 
Here the projector $\pi$ and the inclusion $\iota$ are defined by $\pi
\widetilde{\psi}=\widetilde{\psi}|_{\dR^+}$ for $\widetilde{\psi}\in
L^2(\dR,dq)$ and $(\iota \phi)(q)=\phi(q)$ for $q>0$ while $(\iota
\phi)(q)=0$ otherwise for $\phi\in\CH$. The operator $\hat{q}$, whose
spectral family was used to restrict the Hilbert space, remains a
self--adjoint multiplication operator on $\CH$.  But the momentum
operator $\widetilde{\CO}=\hat{p}=-i\hbar d/dq$, commonly used as the
other fundamental observable on $\widetilde{\CH}$, projects down to a
derivative $\CO=-i\hbar d/dq$ on $\CH$, which is no longer
self--adjoint: The domain of definition of $\CO$ defined by the
projection is
$\CD(\CO)=\CD(\widetilde{\CO})\cap\CH=\{\psi\in\CH:\psi\mbox{
  absolutely continuous, }\psi'\in\CH\mbox{ and }\psi(0)=0\}$.  Its
adjoint has, however, the larger domain of definition
$\CD(\CO^*)=\{\psi\in\CH:\psi\mbox{ absolutely continuous and
  }\psi'\in\CH\}$, whereas $\CO^{**}=\CO$. This shows that $\CO$ is
not essentially self--adjoint, and, even worse, it has no
self--adjoint extensions (cf, e.g., Ref.~\cite{Thirr3Aufl2}).

Being a consequence of the boundary, the latter problem can easily be
cured by using the self--adjoint operator $\frac{1}{2}(\hat{q}\hat{p}+
\hat{p}\hat{q})= -i\hbar(q\,d/dq+1/2)$ (as quantization of $qp$)
instead of $\hat{p}$. Due to the presence of $\hat{q}$, its projection
to $\CH$ no longer needs additional boundary conditions to be
hermitean, and it can easily be shown to be self--adjoint (it
generates the unitary transformation $\psi(q)\mapsto \sqrt{t}\psi(t
q)$). This is related to the fact that the flow of $qp$ is tangential
to the boundary.  We are again lead to the same fundamental
observables as when using the group theoretical quantization and we
obtain unitarily equivalent quantum theories.

\subsubsection{Outlook on possible applications} 

The main advantage of the projection quantization 
(within its limited domain of applicability) as opposed to other
quantization schemes is the fact that it makes use of the quantization
of the embedding phase space $\widetilde \CP$. So, parts of the steps
in the transition from the classical to the quantum system are taken
from the auxiliary system $\widetilde \CP$ and need not be repeated
for $\CP$. This will become particularly transparent at the example in
the subsequent section.

At the level of symplectic manifolds there is a related method, known
as ``symplectic cutting'' in the mathematical literature (cf e.g.\ 
Ref.~\cite{scuts}).  In this approach one is given a torus action with
momentum map on a phase space $\widetilde \CP$. By means of this
momentum map $\widetilde \CP$ can be cut into pieces one of which is
determined by setting the Hamiltonians of the torus action greater
than zero yielding a certain compactification of the subspace $\CP$
defined above. Note that due to the abelian character of the torus
$U(1)^n$ these Hamiltonians always Poisson commute. In Ref.\ 
\cite{scutproj} this technique is employed to prove that the
projection quantization yields a
correct quantization of $\CP$ for a large class of systems. 

Far more complicated examples for the projection quantization than those
provided in this paper, for which the procedure can be relevant (when
extended appropriately to deal with constrained systems), are given by
gravitational theories. 
There, the (symmetric) matrix $g_{\mu\nu}$ of
the coefficients of the metric $g$ in some local chart is required to
satisfy det$(g)\neq 0$ (for all points of spacetime) or, more
precisely, $\pm \mbox{det}(g) >0$, the sign depending on the signature
of the metric $g$. E.g., in a dreibein formulation of Hamiltonian
general relativity one has to require $\det e>0$ for the dreibein
components in order to extract the nondegenerate sector.  In the
context of lattice quantum gravity the implementation of this
condition at the quantum level, as compatible with the general
projection quantization 
above, has been investigated in Ref.~\cite{Loll}.

We conclude these considerations with a cautionary remark: In the
context of gravity theories --- but also, more generally, of
constrained Hamiltonian systems with (additional) ``disallowed
regions'' in phase space --- further care is needed when considering
projection quantization 
(in addition to the standard problems of the
quantization of constrained systems). This becomes obvious already
classically: First removing disallowed regions from phase space
(degenerate sectors in gravity theories) and then performing the
symplectic reduction is in general only equivalent to first reducing
and then singling out the disallowed equivalence classes (or the
equivalence classes without an allowed representative) if the flow of
the constraints does not connect allowed with disallowed regions.

This condition is violated in several popular formulations of gravity
theories in spacetime dimensions four (Ashtekar formulation), three
(Chern--Simons formulation), and two ($BF$-- or, more generally,
Poisson Sigma formulation). In all of these cases, equivalence with
the original, metrical formulation can be established only on the {\em
  nondegenerate\/} sector of phase space and (in contrast to the
original diffeomorphism constraints) the flow of the constraints in
the new formulation does indeed enter the degenerate sector.

To show that this can be of relevance, we provide a simple example (cf
also Ref.~\cite{Peter} for a similar illustration): Consider a particle in
$\dR^3$ with the (original, first class) constraint $C=x \, [(x+2)^2-
(p_x)^2-1] \approx 0$, declaring the subspace with $x \le 0$ to be
``disallowed'' (``degenerate sector'').  Clearly, the flow of $C$ does
not leave (or enter) the forbidden region in phase space. Thus,
removing the disallowed subspace and performing the symplectic
reduction commute, leading to a reduced phase space (RPS) which is a
{\em two--fold\/} covering of $T^* \dR^2$. On the other hand, within
the allowed region of the original phase space the constraint $C$ may
be replaced equivalently by $\widetilde C \equiv (x+2)^2- (p_x)^2-1
\approx 0$. However, the above condition on the flow of the constraint
is no more satisfied in this case. Indeed, while certainly one obtains
the same RPS as before when one first removes the disallowed region
and only then performs the symplectic reduction (which requires
knowledge about the global topology of the orbits), the
(simpler) symplectic reduction of the original theory $T^* \dR^3$ with
respect to $\widetilde C$ leads to only a {\em single\/} copy of $T^*
\dR^2$ as RPS (each point of which contains allowed representatives).

Accordingly, given a procedure for solving the constraint of the original
system (defined in $T^* \dR^3$) at the quantum level, it will yield
inequivalent results when performed with respect to the constraints
$C$ and $\widetilde C$, even if in a second step projection quantization 
(adapted appropriately to the context) is applied to take care of
$x>0$.

Explicit examples of gravity theories in two \cite{Peter} and
three \cite{Matschull} spacetime dimensions showed that the above
mechanism can indeed produce inequivalent factor spaces and,
accordingly, also quantum theories. --- Note, however, that in
this context the failure of projection quantization 
does not result from its insufficiency as a quantization scheme;
rather, the deficiency is evident already on the classical level and
results from the reformulation of the constraints, equivalence in
nondegenerate sectors being, in this context, insufficient for full
equivalence.

\section{The phase space $\CS=S^1 \times \dR^+$}
\label{s:S}

In this section we present the quantization of the phase space $\CS$
which is the restriction of the cotangent bundle $T^* S^1 \sim S^1
\times \dR$ with canonical symplectic form $\omega = d \varphi \wedge dp$
to positive values of the momentum variable $p$.  We denote this
restriction by $S^1 \times \dR^+$, in analogy to $T^* \dR^+\sim \dR^+
\times \dR$.

As stressed already in the previous section, $T^* \dR^+$ is
symplectomorphic to $T^* \dR$; as a symplectic manifold there is {\em
  no}\/ difference between the phase spaces $T^* \dR^+$ and $T^* \dR$
(there is only a difference between what we call the {\em physical}\/
momentum and position).  {\em Topologically}\/ we certainly also have
$\CS \sim S^1 \times \dR$. So we may ask if possibly $\CS$ is also
symplectomorphic to $T^* S^1$. If this were the case, the quantization
of $\CS$ would be immediate, as then we could use the quantum theory
of $T^* S^1$, recapitulated in the previous section.

In contrast to $T^* \dR$ and $T^* \dR^+$, $S^1 \times \dR = T^* S^1$
and $S^1 \times \dR^+ = \CS$ are in fact {\em not}\/ symplectomorphic.
This may be proved by the following simple consideration: Suppose they
were symplectomorphic. Then the diffeomorphism between the two phase
spaces has to map a noncontractible, nonselfintersecting loop on $\CS$
to a likewise loop on $T^* S^1$. Each of these loops separates the
respective phase space into two disconnected parts. On $\CS=S^1 \times
\dR^+$ {\em one}\/ of these two parts has a {\em finite}\/ symplectic
volume. Its image on $T^* S^1$ under the diffeomorphism has an
infinite symplectic volume, on the other hand.  This is in
contradiction with a symplectomorphism, which leaves symplectic
volumes unchanged.

\subsection{$SO^{\uparrow}(1,2)$ and its action}
\label{s:so}

Thus $\CS$ cannot be a cotangent bundle.  However, $\CS$ is the
restriction of a cotangent bundle over $S^1$ to positive values of the
canonical momentum. Such spaces were considered in Subsec.\ \ref{s:sub}
(called $\CP_+$ there).  We thus may apply those considerations to
construct a transitive, almost effective, and canonical group action
on $\CS$.  In particular, as a consequence of Lemma \ref{CS0}, it is
only necessary to find a (finite--dimensional) subgroup of Diff$(S^1)$
with a lift acting transitively on $\CS$. Its action will then be also
effective and have a momentum map.

\subsubsection{Finite-dimensional subgroups of {\rm \Diff($S^1$)}
with transitive action\\ on $\CS \subset T^* S^1$}

The Lie algebra diff($S^1$) of Diff($S^1$) may be represented by
vector fields of the form $v=f(\varphi) \, d/d \varphi$, where $f$ is
a $2 \pi$--periodic function. Thus a dense subalgebra of diff($S^1$)
is spanned by
\begin{equation}
  T=\frac{d}{d\varphi}\quad ,\quad S_k=\sin (k \phi) \, 
   \frac{d}{d\varphi}\:\mbox{
    and }\: C_k=\cos (k \phi) \, \frac{d}{d\varphi} \:\mbox{ with }\: k \in
    \dN \equiv \{1,2,\ldots\} \, ,
\end{equation}
and we will denote it as
$$
\diff_0(S^1):=\{b_0T+\sum_{k>0}(b_kC_k+b_{-k}S_k)|b_k\in\dR\mbox{
  and }b_k= 0\mbox{ for almost all }k\}.
$$

As already mentioned, we are interested in {\em finite--dimensional\/}
subgroups of the diffeomorphism group. They can have an arbitrary
dimension as the following construction shows: To any $n\in\dN$ we can
choose $n$ vector fields on the circle which have disjoint compact
supports. They generate the $n$--dimensional abelian subgroup $\dR^n$.
Clearly, these subgroups have fixed points and thus do not act
transitively on $S^1$ (and neither do their lifts to $\CS$).

To eliminate these and similar subgroups from our consideration we
will, in the following, constrain ourselves to the (still
infinite--dimensional) subalgebra $\diff_0(S^1)$ of $\diff(S^1)$
generated by finite linear combinations of $T$, $S_k$ and $C_k$ in
some chart of $S^1$.  As subalgebras of $\diff(S^1)$ they depend on
the coordinate on $S^1$: Subalgebras corresponding to different
coordinates are not identical; however, they are conjugate to one
another and are thus isomorphic.  The restriction to $\diff_0(S^1)$
will allow us to draw much stronger conclusions, namely we will find
that {\em all\/} finite--dimensional subgroups of $\Diff_+(S^1)$ (the
component of $\Diff(S^1)$ connected to the identity) with Lie algebra
lying in $\diff_0(S^1)$ and with transitively acting lift to $\CS$ are
covering groups of $SO^{\uparrow}(1,2)$:

\begin{theo}\label{unique}
  Each finite--dimensional subgroup of $\Diff_+(S^1)$ which is
  generated by finite linear combinations of $T$, $S_k$ and $C_k$ in
  some chart of $S^1$ and which has a transitively acting lift to
  ${\cal S}\subset T^{\star}S^1$ is isomorphic to a covering group of
  $SO^{\uparrow}(1,2)$ (the $l$--fold covering being generated by
  $l^{-1}T$, $l^{-1}S_l$, and $l^{-1}C_l$).
\end{theo}

\subsubsection{Finite--dimensional subalgebras of the
  Witt algebra}

To prove the theorem we first consider finite--dimensional subalgebras
of the complexification of $\diff_0(S^1)$, which is known as the Witt
algebra
$$
{\cal W}:=\{\sum_{k\in\dZ}a_kL_k|a_k\in\dC\mbox{ and }a_k=0\mbox{
  for almost all }k\}
$$
with generators $L_k = -i \exp(ik\phi) d/d\phi$, $k \in
\dZ$ and relations $[L_j,L_k]=(k-j) L_{k+j}$.

\begin{lemma}\label{Witt}
  The finite--dimensional subalgebras of\/ ${\cal W}$ are at most (complex)
  three--dimen\-sional, in which case they are isomorphic to $sl(2,\dC)$.
\end{lemma}

\begin{proof}
  Let $A$ be a finite--dimensional subalgebra of ${\cal W}$ with at
  least three generators.  Without any restriction these generators
  can be assumed to be of the form $g=L_-+cL_0+L_+$ with
  $L_-=\sum_{k=1}^{M_-}a_kL_{-k}$, $L_+=\sum_{k=1}^{M_+}b_kL_k$,
  $c,a_i,b_i\in\dC$ and $a_{M_-}\not=0\not=b_{M_+}$. Otherwise they
  can be brought into this form by appropriate linear combinations.
  Let $g_i=L_-^{(i)}+c^{(i)}L_0+L_+^{(i)}$, $i=1,2$, be two of the
  generators and $M_{+/-}^{(i)}$ as defined above. Then we can reveal
  the following conditions for $A$ to be finite--dimensional:

  (i) $M_+^{(1)}=M_+^{(2)}$, and analogously $M_-^{(1)}=M_-^{(2)}$:
  Otherwise $[g_1,g_2]$ would contain a contribution of
  $L_{M_+^{(1)}+M_+^{(2)}}$ with nonzero coefficient. By induction,
  repeated commutators would contain contributions from
  $L_{mM_+^{(1)}+nM_+^{(2)}}$ with arbitrary $m,n\in\dN$. Therefore,
  the subalgebra could not be finite--dimensional.

  (ii) $L_+^{(1)}\propto L_+^{(2)}$, and analogously $L_-^{(1)}\propto
  L_-^{(2)}$: Otherwise by appropriate linear combinations we could
  trade the generators for two new generators not fulfilling condition
  (i).

  We conclude that all generators are of the form
  $g_i=a_iL_-+c_iL_0+b_iL_+$, i.e., there are only three linearly
  independent generators $\{L_-,L_0,L_+\}$. This proves that a
  finite--dimensional subalgebra is at most three--dimensional.

  We can now determine the form of these subalgebras $\langle
  L_-,L_0,L_+\rangle$: From the commutation relations of the $L_k$ 
  it follows that $[L_0,L_+]$ has to be proportional to $L_+$ in order
  for $\langle L_-,L_0,L_+\rangle$ to be closed under commutation.
  This can only be the case if there is an $l\in\dN$ such that
  $L_+\propto L_l$. Analogously there must be a $j\in\dN$ such that
  $L_-\propto L_{-j}$. Now we must have $l=j$ because otherwise
  $\langle L_-,L_0,L_+\rangle$ would not be closed.  The only
  three--dimensional subalgebras of ${\cal W}$ are, therefore, given
  by $\langle L_{-l},L_0,L_l\rangle$ for $l\in\dN$, which are easily
  seen to be isomorphic to $sl(2,\dC)$.
\end{proof}

Thus we know all three--dimensional subalgebras of the Witt algebra.
We will see now that they also include all two--dimensional
subalgebras:

\begin{lemma}\label{twodim}
  Each (complex) two--dimensional subalgebra of\/ ${\cal W}$ is a subalgebra
  of one of the $sl(2,\dC)$ subalgebras found in the preceding lemma.
\end{lemma}

\begin{proof}
  A two--dimensional Lie algebra generated by $g_1$ and $g_2$ can,
  without restriction, be assumed to be of the form $[g_1,g_2]=0$ or
  $[g_1,g_2]=g_2$, respectively. In close analogy to the proof of the
  preceding lemma, one may show that the former case implies $g_1
  \propto g_2$ (in contradiction to the linear independence of $g_1$
  and $g_2$) and that the latter case is possible only if $g_1\propto
  L_0$ and $g_2\propto L_l$ for an $l\in\dZ$.
\end{proof}

To use the information about ${\cal W}$ contained in the preceding
two lemmas we have to translate it to the real form $\diff_0(S^1)$.
The statements on (now real) dimensionality in Lemma~\ref{Witt} and
Lemma~\ref{twodim} remain true because otherwise we could construct
contradictions to these lemmas by complexification.

\subsubsection{$SO^{\uparrow}(1,2)$ and its covering groups}

The $sl(2,\dC)$--subalgebras $\langle L_{-l},L_0,L_l\rangle<{\cal
  W}$ have the real forms $\langle l^{-1}T,l^{-1}S_l,l^{-1}C_l
\rangle$ as subalgebras of $\diff_0(S^1)$. For any $l$ this is an
  $so(1,2)$--algebra shown by the isomorphism
 \be \frac{T}{l} \leftrightarrow T_0
\, , \; \, \frac{S_l}{l}\leftrightarrow T_1 \, , \; \, 
\frac{C_l}{l} \leftrightarrow T_2 \,.
\label{Lieiso1} \ee Here the
$T_i$, $i=0,1,2$ are generators of $so(1,2)$, satisfying the standard
relations $[T_i,T_j]=\varepsilon_{ij}{}^k \, T_k$, where
$\varepsilon_{012} = 1$ and indices are raised by means of
diag$(-1,1,1)=\kappa/2$, $\kappa$ being the Killing metric. As real forms of
$sl(2,\dC)$ the above subalgebras are unique by demanding them to be
real subalgebras of the real form $\diff_0(S^1)$ of ${\cal W}$.

For later use it is worthwhile to exploit the Lie algebra isomorphisms
of $so(1,2)$ to $sl(2,\dR)$ and $su(1,1)$. An isomorphism between the
former two in terms of their generators $T_i$ and $\sigma_+$, $\sigma_-$,
$\sigma_3/2$, respectively, is given by \be
T_0\leftrightarrow\frac{1}{2}(\sigma_+-\sigma_-)\quad, \qquad
T_1\leftrightarrow\frac{1}{2}(\sigma_++\sigma_-)\quad, \qquad 
T_2\leftrightarrow\frac{1}{2}\sigma_3 ,
\label{Lieiso2} \ee  where $2 \, \sigma_\pm = \sigma_1 \pm i \sigma_2$ 
and $\sigma_j$,
$j=1,2,3$ denote the standard Pauli matrices.  An isomorphism between
$so(1,2)$ and $su(1,1)$ is provided by 
\be T_0\leftrightarrow -\frac{i}{2}\sigma_3 \,
 , \; \, T_1 \leftrightarrow \frac{1}{2}\sigma_1 \, , \; \, T_2 
 \leftrightarrow \frac{1}{2}\sigma_2
 \, . \label{Lieiso3} \ee

The subgroup of $\Diff_+(S^1)$ generated by $l^{-1}T$, $l^{-1}S_l$,
and $l^{-1}C_l$ is the $l$--fold {\em covering\/} group of
$SO^{\uparrow}(1,2)$. This is the case because $\left(\exp(2\pi
  l^{-1}T)\right)^j=\exp(2\pi jl^{-1}d/d\varphi)\not=1$ for $0<j<l$
and $\left(\exp(2\pi l^{-1}T)\right)^l=1$.  In the language of
$SO^{\uparrow}(1,2)$ ($l=1$), $T$ generates rotations in the
$(x_1,x_2)$--plane of the $(2+1)$--dimensional Minkowski space, and
$S_1$ and $C_1$ generate boosts along the $x_1$-- and
$x_2$--direction, respectively.

We thus arrived at $SO^{\uparrow}(1,2)$ and its covering groups as
maximal finite--dimensional subgroups of $\Diff_+(S^1)$ with Lie
algebra in $\diff_0(S^1)$.  They are maximal finite--dimensional
subgroups of $\Diff_+(S^1)$ in the sense that there is no
finite--dimensional subgroup of $\Diff_+(S^1)$ which has one of these
groups as a subgroup.  This follows easily from the fact that their
complexified Lie algebras contain the element $L_0$.

For Theorem~\ref{unique} to hold the restriction to $\diff_0(S^1)$ is
essential: As already noted at the beginning of this subsection,
$\Diff_+(S^1)$ contains finite--dimensional subgroups of arbitrary
dimension. The examples provided there were of no interest in our
context, however; for physical applications, moreover, it seems
natural to restrict oneself to finite linear combinations of
trigonometric functions as for the {\em fundamental\/} observables
(certainly this does not imply that all the observables are restricted
in the same manner, since for them one still is allowed to take
infinite linear combinations, cf also Subsec.\ \ref{Generating}
above).

Finally, to prove Theorem~\ref{unique} we are left to study their
possible subgroups.

\begin{lemma}\label{subgroups}
  For any covering group of $SO^{\uparrow}(1,2)$ there are two
  conjugacy classes of two--dimensional subgroups (both of which are
  isomorphic to $\dR\semidir\dR$). The Lie algebras of respective
  representatives are spanned by $T_2$ and $T_0 \pm T_1$.
\end{lemma}

\begin{proof}
  As abelian subalgebras of $sl(2,\dR)$ are at most one--dimensional
  ($sl(2,\dR)$ has rank one), any two--dimensional subalgebra may be
  spanned by generators $\tau_+$ and $\tau_3$ satisfying
  $[\tau_3,\tau_+]=\tau_+$. In the complexified Lie algebra
  $sl(2,\dC)$ they span a Borel subalgebra, which is a maximally
  solvable subalgebra and unique up to conjugation. Thus we know that
  for any two--dimensional subalgebra of $sl(2,\dR)$ there is, in the
  fundamental representation of the algebra, a {\em complex\/}
  two--by--two matrix $M$ of unit determinant such that $\tau_3 = M \,
  (\sigma_3/2) \, M^{-1}$ and $\tau_+ = M \, \sigma_+ \, M^{-1}$ (and,
  up to a sign, $M$ is unique). Reality of the matrices $\tau_3$ and
  $\tau_+$ implies that $M$ is either real or purely imaginary. In the
  former case, $M \in SL(2,\dR)$ and the conjugation is compatible
  with the reality condition leading from $sl(2,\dC)$ to $sl(2,\dR)$.
  In the latter case, $M= \widetilde M i \sigma_1$ where $\widetilde M
  \in SL(2,\dR)$. Conjugation with the imaginary piece $i \sigma_1$
  maps $(\sigma_3,\sigma_+)$ into $(-\sigma_3,\sigma_-)$. The
  assertion of the lemma then follows upon the isomorphism
  (\ref{Lieiso2}) and exponentiation to group level.
\end{proof}

To discuss transitivity of group actions on $\CS\subset T^{\star}S^1$,
we finally need the lifts of the diffeomorphisms generated by $T$,
$S_l$, and $C_l$.  According to Eq.\ (5) and the remarks in Subsec.\ 
\ref{s:general}, they are generated by the Hamiltonian vector fields
\be T \to \{ \cdot,p \} \, , \; \, S_l \to \{\cdot, p \, \sin l
\varphi\} \, , \; \, C_l \to \{\cdot, p \, \cos l \varphi \},
\label{moment}\ee respectively. This also provides a momentum map for the
action of $SO^{\uparrow}(1,2)$ on $\CS$.

We are now in the position to prove our theorem:

\begin{proof}[(Of Theorem \ref{unique})]
  According to Lemma 4 the finite--dimensional subgroups of
  $\Diff_+(S^1)$ which are generated by elements of $\diff_0(S^1)$ can
  be at most three--dimensional because the finite-dimensional
  subalgebras of ${\cal W}$, which is the complexification of
  $\diff_0(S^1)$, are at most three-dimensional.

  The three-dimensional of these subgroups are isomorphic to $l$--fold
  covering groups of $SO^{\uparrow}(1,2)$ spanned by $l^{-1}T$,
  $l^{-1}S_l$, and $l^{-1}C_l$. All the two--dimensional subgroups are
  subgroups of these three--dimensional ones, moreover. Finally, there
  are the one--dimensional subgroups of $\Diff_+(S^1)$ which are
  generated by exponentiation of an arbitrary element of
  $\diff_0(S^1)$.  We now investigate the action of these subgroups
  when lifted to $\CS \subset T^*S^1$.

  One--dimensional groups cannot have orbits filling all of the
  two--dimensional half--cylinder. So they cannot act transitively.

  According to Lemma~\ref{subgroups} and Eq.\ (\ref{Lieiso1}), all
  two--dimensional subgroups are in one of the two conjugacy classes,
  representatives of which are generated by the vector fields $C_l$
  and $T \pm S_l$. Their lifts $\{\cdot, p\cos l\varphi \}$ and
  $\{\cdot , p(1\pm\sin l\varphi) \}$ to $\CS$ fix the fiber over
  $\varphi=\mp \pi/(2l)$ and therefore the groups cannot act
  transitively. (The other two--dimensional subgroups, being conjugate
  to one of these two groups, can just as less act transitively.)

  The only candidates with transitively acting lift are now the
  covering groups of $SO^{\uparrow}(1,2)$. That they act indeed
  transitively can be seen from the following consideration: The lift
  of the action of an $l$--fold covering group of $SO^{\uparrow}(1,2)$
  is generated by the two vector fields given in the previous
  paragraph together with the vector field $\{\cdot, p\}$. The former
  two act transitively in some fibers and the latter one acts fiber
  transitively.  Thus their joint action is transitive on ${\cal S}$.
\end{proof}

\subsubsection{Integrating the group actions}
\label{integrating}
In this subsection we will derive the finite action on $\CS$ generated
by $T$, $S_l$, and $C_l$. There is a well known
$SO^{\uparrow}(1,2)$--action on $S^1$ given by (see, e.g.,
Ref.~\cite{Olive}) \be z\mapsto\frac{\alpha
  z+\beta}{\overline{\beta}z+\overline{\alpha}}\quad,\qquad z=\exp
i\varphi\in S^1\quad,\qquad A \equiv \left(\begin{array}{cc}\alpha & \beta \\
    \overline{\beta} & \overline{\alpha}\end{array}\right)\in SU(1,1)
\, , \label{wirkung} \ee with $|\alpha|^2-|\beta|^2=1$. This action on
$S^1$ has been written down in terms of $SU(1,1)$, which is a
two--fold covering of $SO^{\uparrow}(1,2)$ (note that $A \in SU(1,1)$
and $-A$ have the same action).

(The relation between these two groups can be made explicit by means
of the action $X\mapsto A X A^\dagger$ of $A \in SU(1,1)$ on matrices
$X=X^\dagger$ satisfying $\mbox{tr}(\sigma_3 X)=0$, where $\dagger$
denotes transposition of the complex conjugate matrix. This
transformation preserves the determinant of $X$, which, in the
parametrization \be X=\left(\begin{array}{cc} x_0 & x_1-ix_2\\x_1+ix_2
    & x_0\end{array}\right)\quad , \qquad x_0,x_1,x_2\in\dR \, ,
\label{X}\ee is nothing but the bilinear form $x_0^2-x_1^2-x_2^2$; in
this way $A$ is seen to generate a (proper) Lorentz (or
$SO^\uparrow(1,2)$) transformation on the $(2+1)$--dimensional
Minkowski space spanned by $(x_0,x_1,x_2)$.)

It is straightforward to verify that the infinitesimal form of
Eq.~(\ref{wirkung}) coincides with the action generated by the vector
fields $T$, $S_1$, and $C_1$ (cf Eqs.\ (5) and (6)). In this way we
may also determine the lift of the (finite) action (\ref{wirkung}) to
$\CS$ (cf Subsec.\ \ref{s:general}), yielding $p \mapsto p|\alpha
e^{i\varphi}+\beta|^2$ in this case.

For $l>1$ the action (\ref{wirkung}) on $S^1$ can be generalized by
substituting $\exp il\varphi$ for $z=\exp i\varphi$. Infinitesimally,
this action is readily seen to coincide with the one generated by the
vector fields $T$, $S_l$, and $C_l$. However, taking the $l$--th root
in a continuous manner to arrive at an action on $\varphi\in \dR
\mbox{ mod } 2\pi$ is nontrivial; in particular, for $l>2$ it does
{\em not} lead to an action on $S^1$ of the group $SU(1,1)$ itself,
but of appropriate covering groups only (contrary to what is
claimed, e.g., in Ref.~\cite{Olive}).

Actually, from the discussion preceding Lemma~\ref{subgroups}, we
already know that the group generated by $T$, $S_l$, and $C_l$ is an
$l$--fold covering group of $SO^{\uparrow}(1,2)$. Thus, we can see
that it is not possible to express the action in terms of $SU(1,1)$
for $l>2$.  Introducing the parameters $\gamma:=\alpha^{-1}\beta$,
$|\gamma|<1$ and $0\leq\omega<2\pi$ by $\alpha=|\alpha|\exp i\omega$
of $SU(1,1)$ (see, e.g., Ref.~\cite{Bargmann}), which make explicit
the topology of $SU(1,1)$, the $n$--fold covering group of
$SO^{\uparrow}(1,2)$ can be parameterized by these parameters taking,
however, $\omega$ in the range $0\leq\omega<n\pi$.  The action
(\ref{wirkung}) on $S^1$ with $\varphi$ replaced by $l \varphi$ now
takes the form \be \exp il\varphi\mapsto\exp
(2i\omega)\frac{\gamma+\exp il\varphi}{\overline{\gamma}\exp
  il\varphi+1} \, .
\label{mapS1} \ee
Acting on $\exp il\varphi$,
$0\leq\varphi<2\pi l^{-1}$, this action on $S^1$ is an almost
effective action of the $n$--fold covering group of
$SO^{\uparrow}(1,2)$, and $n$ does not need to be identical to $l$.
However, to obtain an action on $S^1$, $\varphi$ has to take values in
$[0,2\pi)$ where $\varphi$ and $\varphi+2\pi l^{-1}$ are not to be
identified.

This observation will fix the covering group which acts effectively on
$\dR\mbox{ mod } 2\pi$ if we take the $l$--th root. To this end it
suffices to consider the action for $\gamma=0$ because $\gamma$ takes
values in a simply connected domain. The action reduces to
$$
\exp il\varphi\mapsto\exp(2i\omega)\exp il\varphi
$$
which leads to $\varphi\mapsto\varphi+2l^{-1}\omega$ if we use
continuity and the fact that we have to obtain the identity
transformation for $\omega=0$. The last two conditions fix the branch
of the $l$--th root uniquely. Now $\omega=n\pi$ must give the same
result as $\omega=0$ because we consider the action of the $n$--fold
covering group of $SO^{\uparrow}(1,2)$.  This is possible only if $n$
is an integer multiple of $l$ and an effective action is obtained for
$n=l$. Thus we see that Eq.~(\ref{mapS1}) determines an effective action
of the $l$--fold covering group of $SO^{\uparrow}(1,2)$ on $S^1$ (and
an almost effective action of the $l m$--fold covering for any $m
\in\dN$), but (for $l > 2$) not an $SU(1,1)$--action.

So, following the strategy for finding a group action on $\CS$ as
formulated in Subsec.\ \ref{s:sub}, we thus arrive at the following
action of finite dimensional groups on $\CS \subset T^*S^1$:
\be\label{map2} (\exp il\varphi,p) \mapsto \left(\frac{\alpha\exp
    il\varphi+\beta}{\overline{\beta}\exp il\varphi
    +\overline{\alpha}}, p|\alpha e^{il\varphi}+\beta|^2\right) \, .
\ee This is the lift of the action of the $l$--fold covering group of
$SO^{\uparrow}(1,2)$ presented above. (We were searching for subgroups
of $\Diff(S^1)$ which all act effectively; therefore, these subgroups
are $l$--fold coverings and not $lm$--fold ones ($m>1$).) By
construction, for any $l \in \dN$, Eq.\ (\ref{map2}) provides a
transitive, effective, and Hamiltonian action on $\CS$ with momentum
map.

Two more remarks: First, by means of the above action for $l=2$ we may
identify the phase space $\CS$ with the coset space $SU(1,1)/N$. Here
$N$ denotes the nilpotent subgroup appearing in the Iwasawa
decomposition of $SU(1,1)$ (obtained by exponentiating $T_2-T_0$ in
Eq.~(\ref{Lieiso2}), cf also Ref.~\cite{Schramm} for details). $N$ is
the stabilizer group of any point $(\varphi =0,p) \in \CS$, since
obviously its generator $\{\cdot, p (\cos 2\varphi-1)\}$ vanishes
identically on the fiber over $\varphi=0$.

Second, the (finite) action (\ref{map2}) on $\CS$ may be also obtained
as the (effective) action of the $l$--fold covering of
$SO^{\uparrow}(1,2)$ on the $l$--fold covering of the future light
cone $\CC^+$ in $(2+1)$--Minkowski space $(x_0,x_1,x_2)$. For $l=1$
this is just the fundamental (defining) action of the (proper) Lorentz
group which clearly maps the future light cone ${\cal C}^+\colon
x_0^2-x_1^2-x_2^2=0$, $x_0>0$, onto itself so that its action on
Minkowski space can be restricted to an action on ${\cal C}^+$. The
action (\ref{map2}) with $l=1$ is then obtained from the
$SO^{\uparrow}(1,2)$--action on ${\cal C}^+$ upon identifying $x_0$
with $p$ and the polar angle of the light cone with $\varphi$.  For
$l>1$ this generalizes to \be\label{lightconek}
(x_0,x_1+ix_2)\leftrightarrow(p,pe^{-il\varphi})\quad,\quad
0\leq\varphi< 2\pi, p>0\, ,\ee identifying the phase space $\CS$ with
an $l$--fold covering of ${\cal C}^+$. To verify the equivalence of
the actions one only needs to check the infinitesimal correspondence
(\ref{Lieiso1}) and (\ref{moment}) (with $T_i\in so(1,2)$ interpreted
as the generators on (the $l$--fold covering of) $\CC^+$). Formula
(\ref{map2}) may now be obtained also by this approach via $X\mapsto A
X A^\dagger$, where $A \in SU(1,1)$ as above and $X$ results from
combining Eqs.\ (\ref{X}) and (\ref{lightconek}).

\subsubsection{Admissible group actions on $\CS$}
\label{s:admiss}

We have now determined all the lifts of actions of the subgroups of
$\Diff_+(S^1)$ found in Theorem~\ref{unique}. The possible effectively
acting groups are the $l$--fold covering groups of $SO^{\uparrow}(1,2)$.
We are now left only with checking the validity of the SGP for
these group actions.

As is obvious from Eq.\ (\ref{moment}), the SGP is violated for $l\neq
1$. Alternatively we may also apply Lemma~\ref{generating} due to the
semisimplicity of $SO^{\uparrow}(1,2)$ and its covering groups: In
the case of Eq.~(\ref{map2})  $G$ is identified with the $l$--fold covering
group of  $SO^{\uparrow}(1,2)$, which has a trivial center only for
$l=1$.

In the present case the use of the Lemma was not essential. However,
let us remark that this may change drastically when more complicated
phase spaces and group actions are considered (and in particular for
infinite--dimensional phase spaces).

This fixes the parameter $l$ in the countable family of (effective)
group actions to be $l=1$ so that we end up with a unique effective
action of the group $SO^{\uparrow}(1,2)$.

Any covering group of $SO^{\uparrow}(1,2)$ is, however, allowed as
{\em almost}\/ effectively acting group provided its action projects
down to the $SO^{\uparrow}(1,2)$--action (Eq.\ (\ref{map2}) with
$l=1$).  The most general almost effective action is provided by the
universal covering group $\widetilde{SO}\mbox{}^{\uparrow}(1,2)$ of
$SO^{\uparrow}(1,2)$. According to the considerations of Subsec.\ 
\ref{s:group} we will thus examine the unitary representations of
$\widetilde{SO}\mbox{}^{\uparrow}(1,2)$ for possible quantum
realizations of $\CS$ in the following subsection (using the momentum
map (\ref{moment}) with $l=1$ only).

\subsection{The quantum theory}

In the present subsection we will apply two methods to quantize $\CS$.
The first one completes the group theoretical quantization by using
the group action derived in the preceding subsection. The second
approach employs the projection quantization 
of Subsec.\ \ref{s:restriction}
making use of the fact that $\CS$ is the restriction of $T^*S^1$ to
positive momentum. Quantizing this phase space $\CS$ thereby provides
another example for the application of this method with, in contrast to
$T^*\dR^+$, a discrete spectrum of the observable $\hat{p}$ used to
project down to the restricted Hilbert space. We will find that both
quantization procedures are compatible, and that demanding equivalence
constrains the quantum realizations obtained within group theoretical
quantization.

\subsubsection{Group theoretical quantization of $\CS$}
\label{s:GroupQuant}

According to the results of the previous subsection, when applying the
group theoretical quantization scheme to $\CS$, we are to analyse the
(weakly continuous) unitary IRREPs of the universal covering group
$\widetilde{SO}\mbox{}^{\uparrow}(1,2)$ of $SO^{\uparrow}(1,2)$.

Thus we first have to look for unitary representations of $so(1,2)$.
Its generators $T_0$, $T_1$, and $T_2$ obey the relations
$[T_0,T_1]=T_2$, $[T_0,T_2]=-T_1$ and $[T_1,T_2]=-T_0$. This rank one
algebra has the Casimir operator $C:=T_0^2-T_1^2-T_2^2$. As maximal
set of commuting algebra elements we choose $\{-iT_0,C\}$, which will
be promoted to the maximal set $\{H,C\}$ of commuting operators on a
representation space.

The states in irreducible representations can be classified by the
eigenvalues $\lambda$ and $q$ of $H$ and $C$, respectively. In each
irreducible representation, $T_+:=T_1-iT_2$ and $T_-:=-T_1-iT_2$ act
as raising and lowering operators, respectively, which can be read off
from the relations
$$
[H,T_+]=T_+\quad,\quad[H,T_-]=-T_-\quad,\quad[T_+,T_-]=-2H.
$$
On an orthonormal basis $\{\phi^q_{\lambda}\}_{\lambda\in\Lambda}$
of a representation characterized by the eigenvalue $q$ of $C$ and
indexed by the eigenvalues of $H$, the action of $H$, $T_+$, and $T_-$
is given by \be\label{creation}
H\phi^q_{\lambda}=\lambda\phi^q_{\lambda}\quad,\quad
T_+\phi^q_{\lambda} =
\omega_{\lambda+1}\sqrt{q+\lambda(\lambda+1)}\phi^q_{\lambda+1}\quad,\quad
T_-\phi^q_{\lambda} =
\overline{\omega}_{\lambda}\sqrt{q+\lambda(\lambda-1)}\phi^q_{\lambda-1}
\ee with arbitrary phase factors $\omega_{\lambda}$, which can be
chosen to be $1$ by a unitary change of the basis.  One can see that
the spectra of $H$ in all irreducible representations are
equidistantly spaced by $1$.

A more detailed analysis \cite{Bargmann,Pukansky,Sally} of the
irreducible unitary representations of
$\widetilde{SO}\mbox{}^\uparrow(1,2)$ reveals that there are ---
besides the trivial representation --- the following three families:

\medskip

\begin{tabular}{r|c|c|c}
  & irred.\ rep.\ & $\Lambda$ & $C$\\\hline
 continuous series & $C^{k,q},0\leq k<1, q>k(1-k)$ & $k+\dZ$ & $q$\\\hline
 discrete series & $D^{-k},k\in\dR^+$ & $-k-\dN_0$ & $k(1-k)$\\
  & $D^k,k\in\dR^+$ & $k+\dN_0$ & $k(1-k)$
\end{tabular}

\medskip

We now have to select the appropriate representations from the
mathematically possible ones in accordance with the general principles
outlined in Subsec.\ \ref{s:group}. This will be done by checking the
classical property $p>0$ (in complete analogy with $q>0$ for
$T^*\dR^+$, cf Subsec.\ \ref{s:TRp}). According to Eqs.\ 
(\ref{Lieiso1}) and (\ref{moment}) this enforces the spectrum
$\Lambda$ of $H$ to be purely positive.

In the continuous series
the spectrum $\Lambda$ is unbounded from both sides so that these
representations are to be disregarded. The same applies to the
negative discrete series, where the spectrum is purely negative.
The condition of positive spectrum of $H$ is thus fulfilled only in
the positive discrete series (for arbitrary parameter $k\in\dR^+$).
In this case there is a ground state $\phi^q_{\lambda_0}$,
$\lambda_0=k$, $q=k(1-k)$, which is
annihilated by $T_-$.

The choice $D^k$ now already determines the quantum theory of
$\CS$ in the group theoretical framework. In the following we provide
one possible realization of this Hilbert space by means of
antiholomorphic functions on the unit disc\footnote{For further realizations
  we refer to Ref.~\cite{Schramm} and to Subsec.\ \ref{s:Hardy} below}.

For $k>\frac{1}{2}$ a representation on the Hilbert space ${\cal
  H}_k(\CD)$ of antiholomorphic functions on the unit disc ${\cal
  D}:=\{z\in\dC|z\overline{z}< 1\}$ with inner product \be
(f,g)_k=\frac{2k-1}{2\pi i}\int_{\cal
  D}\overline{f(z)}g(z)(1-z\overline{z})^{2k-2}dzd\overline{z} \ee is
given by \cite{Sally} \be
(D^k(\gamma,\omega)f)(\overline{z})=\exp(2ik\omega) (1-|\gamma|^2)^k
(\overline{\gamma}\overline{z}+\exp(2i\omega))^{-2k}
f\left(
\frac{\overline{z}+\gamma\exp(2i\omega)}{\overline{\gamma}\overline{z}+
\exp(2i\omega)}\right),
\ee
where $(\gamma,\omega)$ parameterize the universal covering of
$SU(1,1)$ (see Subsec.~\ref{integrating}). The factor $\exp(2ik\omega)$
determines for which values of $k$ the representation can be projected
to a representation of $SU(1,1)$ or $SO^{\uparrow}(1,2)$.

An orthonormal basis of ${\cal H}_k(\CD)$ which diagonalizes $H=-iT_0$
is given by the functions \be\label{ONbasis}
g_{k,n}(z)=\sqrt{\frac{\Gamma(2k+n)}{\Gamma(2k)\Gamma(n+1)}} \:
\overline{z}^n\quad,\quad n\in\dN_0.  \ee For $0<k\leq\frac{1}{2}$ the
Hilbert space ${\cal H}_k(\CD)$ can be defined by completing the span
of the orthonormal basis $\{g_{k,n}\}_{n\geq 0}$.

By differentiating and using Eq.~(\ref{Lieiso3}), we get the
representations
\begin{eqnarray}
  H & = & k+\zq\frac{d}{d\zq} \nonumber\\
  T_+ &=& -2k\zq-\zq^2\frac{d}{d\zq} \label{actHsD}\\
  T_- & = & - \frac{d}{d\zq}\nonumber
\end{eqnarray}
of the generators $T_0=iH$, $T_1= (T_+ - T_-)/2$, and $T_2 = i(T_+ +
T_-)/2$ of $SU(1,1)$. On the elements $g_{k,n}$ of the orthonormal basis
(\ref{ONbasis}) they act as
\begin{eqnarray}
  Hg_{k,n} & = & (k+n)g_{k,n}\nonumber\\
  T_+g_{k,n} & = & -\sqrt{(2k+n)(n+1)}g_{k,n+1}    \label{actiongms}\\
  T_-g_{k,n} & = & -\sqrt{n(2k+n-1)}g_{k,n-1}\nonumber
\end{eqnarray}
which is identical to Eqs.~(\ref{creation}) if we use the relations
$\lambda=k+n$ and $q=k(1-k)$, choosing the phases $\omega_{k+n}$ to be
$-1$.

According to Eq.~(\ref{Lieiso1}) the spectrum of $p$ in a quantization
of $\CS$ using the $SO^{\uparrow}(1,2)$ action is given by the
spectrum of $H$.  Reintroducing Planck's constant (cf the discussion
in Subsec.\ \ref{s:quant}), we get the following quantization map
\be\label{quant} \hat{p}=\frac{\hbar}{i}T_0=\hbar
H\quad,\quad\widehat{(p\sin\varphi)} =\frac{\hbar}{i}T_1\quad,
\quad\widehat{(p\cos\varphi)}=\frac{\hbar}{i}T_2.  \ee

Thus we obtain a one--parameter family of inequivalent quantizations
with spectra $\hbar (k+\dN)$, $k\in\dR^+$, of $\hat{p}$. On the other
hand, from the point of view of geometric quantization
\cite{Woodhouse}, the ambiguity in different quantum realizations
should be parameterized by a parameter living on a circle
($\theta$--angle) (cf our discussion in Sec.\ \ref{Standard} and, for
the particular phase space $\CS$, Ref.\ \cite{scutproj}).
Similarly, application of the alternative projection quantization, 
which is presented in the subsequent
subsection, will be seen to yield $k \in (0,1]$ or, better, $k \in
S^1$.

From this we conclude that representations characterized by values of
$k$ larger than one should be regarded as ``unphysical'' in the group
theoretical quantization --- similar to discarding the continuous or
negative discrete series of representations. Note, however, that {\em
  within\/} the scheme of group theoretical quantization this cannot
be obtained by a natural condition such as $p>0$, because all values
of $k$ are obtained here on an equal footing. (Restriction to
representations of effectively acting admissible groups, on the other
hand, leads to $k \in \dN$ only; this merely excludes the
$\theta$--parameter (obtained from permitting also almost effective
group actions) and still leaves $k$ unbounded.)

\subsubsection{Quantum realization via restriction of a Hilbert space}
\label{s:Restrict}

By definition, our phase space $\CS$ is the restriction of $T^*S^1$ to
positive values of the canonical momentum $p$ such that it can be
treated by using projection quantization. 
Quantizing $\CS \subset T^*S^1$, we thus proceed as follows: We {\em
  first}\/ quantize $T^*S^1$, which is standard and which we reviewed
in Sec.\ \ref{s:quant} (from various perspectives). Thereafter, in a
second step, we implement the condition $p>0$ using the projector to
the positive part of the spectrum of $\hat{p}$. (Cf Subsec.\ 
\ref{s:restriction} for the strategy in general context.)

More precisely, in Sec.\ \ref{s:quant} we observed that the spectrum
of $\hat{p}$ in the Hilbert space $\widetilde{\CH}_{\theta}$ spanned
by quasi--periodic functions on $S^1$ characterized by $\theta$ with
inner product $(f,g)=(2\pi)^{-1}\int_{S^1}\overline{f}gd\varphi$ is
$\{ \hbar(m + \theta), m \in \dZ\}$.  The respective eigenstates
$f_{\theta,m}:=\exp(i(m+\theta)\varphi)$, $m \in \dZ$, form an
orthonormal basis of $\widetilde{\CH}_{\theta}$. The condition
$\hat{p}>0$ is met on any subspace $\CH_{\theta+m_{min}}$ of
$\widetilde{\CH}_{\theta}$ which is spanned by the vectors
$f_{\theta,m}$ with $m \ge m_{min} \in \dN_0$.

According to the general strategy of projection quantization 
in Subsec.~\ref{s:restriction} we have to demand here $m_{min}=0$ to
obtain the maximal Hilbert subspace on which $\hat{p}>0$ is fulfilled.
As Hilbert spaces of $\CS$ we will only regard $\CH_{\theta}$, i.e.,
those with $m_{min}=0$. In the case of $T^*\dR^+$ the requirement of
maximality was necessary so as to reproduce standard results on the
quantization of this phase space (including those of group theoretical
quantization).  To achieve maximality also in the case of $\CS$, on
the other hand, forces us to restrict the outcome of the group
theoretical quantization by declaring representations with $k>1$ as
``unphysical''. For mathematical reasons it is, however, instructive
in some contexts to leave $m_{min}$ unspecified and discuss
observables on all spaces $\CH_{\theta+m_{min}}$; we will do so in
Subsecs.\ \ref{s:Equiv} and \ref{s:Hardy} below.

All infinite--dimensional, separable Hilbert spaces are isomorphic to
one another; additional structures arise only through the
representation of some elementary set of observables in
$\CH_{\theta+m_{min}}$, which is induced by the respective
representation in $\widetilde{\CH}_{\theta}$.

We choose $p$ and $U := \exp i \varphi$ as such a set of elementary
functions.  Their action on the basis $\{f_{\theta,m},m \geq
m_{min}\}$ of $\CH_{\theta+m_{min}}$ is provided by
$\hat{p}f_{\theta,m} = \hbar(m+\theta)f_{\theta,m}$ and
$\hat{U}f_{\theta,m}=f_{\theta,m+1}$, where $\hat{U}$ is the obvious
multiplication operator and $\hat{p}=-i\hbar \,(d/d\varphi)$. The
Poisson algebra $\{ U, p\}=iU$ is turned correctly into the
commutation relations $[\hat{U},\hat{p}]=- \hbar \hat{U}$.

Classically $p>0$ and $\overline U \, U \equiv U \, \overline U = 1$.
By construction of $\CH_{\theta+m_{min}}$, $\hat{p}$ becomes positive
also as an operator, and it remains self--adjoint.  On the other hand,
$\hat{U}$ although unitary in $\widetilde{\CH}_{\theta}$, is only
isometric in $\CH_{\theta+m_{min}}$: one still finds $\hat{U}^*
\hat{U} =\dId$, $\hat{U}^*$ denoting the adjoint of $\hat{U}$, but
now, due to the existence of a lowest lying state $f_{\theta,m_{min}}$
in $\CH$ (which can be interpreted as corresponding classically to the
boundary $p=0$ of $\CS$), $\hat{U} \hat{U}^*$ is equal only to the
projector $\dId - P_{m_{min}} \neq \dId$ (where $P_{m_{min}}$ is the
projector on the state $f_{\theta,m_{min}}$).  Such a feature has been
observed already in the general context in Subsec.\ 
\ref{s:restriction}, and one can ask for a substitute of $\hat{U}$
with improved properties.  However, as will be found in the next
subsection, the operator corresponding to $\exp i \varphi$ cannot be
made unitary in the group theoretical approach as well. Since
unitarity of $\hat{U}$, generating translations in $p$ as a
consequence of the commutation relations, is incompatible with the
restriction of the phase space, isometry is the most that can be
achieved for $\hat{U}$ in a quantum theory of $\CS$.

We finally remark that over the complex numbers the Poisson algebra of
$U$ and $p$ is a two--dimensional affine Lie algebra and indeed
$\CH_{\theta+m_{min}}$ provides an irreducible representation of it.
However, this representation is {\em not}\/ unitary (even in
$\widetilde{\CH}_{\theta}$) and it cannot be so as a consequence of the
complex structure constants appearing already in the {\em classical}\/
Poisson algebra.

The classical ($U$,$p$)--algebra closes over the real numbers only
when taking the real and imaginary part of $U$, $\cos \varphi$ and
$\sin \varphi$, as separate generators. Together with $p$ they then
provide the Lie algebra of $E_2$ and this was precisely the algebra
that yielded $\widetilde{\CH}_{\theta}$, the quantum theory for $T^*
S^1$, and not the present quantum realization in
$\CH_{\theta+m_{min}}$.  This mirrors the fact that $\{\cdot, \cos
\varphi\}$ and $\{\cdot, \sin \varphi\}$ cannot be used as generating
vector fields on $\CS \subset T^*S^1$ (being transversal to the
boundary $p=0$), so that they do not exponentiate to the action of a
group on $\CS$. To apply the group theoretical approach we, therefore,
needed to discuss the more involved group actions provided in the
previous subsection.

\subsubsection{Equivalence of the two approaches}
\label{s:Equiv}

If we compare the spectra of $\hat{p}$ obtained in the approaches
above, we see that they are compatible: With the identification
$\theta+m_{min}=k$ of the respective parameters labeling the Hilbert
spaces, the operators $\hat{p}$ of the two quantizations can be
identified. We are thus lead to the following Hilbert space
isomorphism between $\CH_{\theta + m_{min}}$ and $\CH_k(\CD)$:
$f_{\theta,n+m_{min}}\mapsto g_{k,n}$, $n\in\dN_0$.

The identification of the creation operator $\hat{U}$ of
Subsec.~\ref{s:Restrict} with the appropriate operator in
Subsec.~\ref{s:GroupQuant} is somewhat more involved.  Classically,
$U=\cos\varphi+i\sin\varphi$. Thus a first ansatz, ignoring factor
ordering problems, for defining the operator $\hat{U}$ in
Subsec.~\ref{s:GroupQuant} could be of the form $T_+H^{-1}$, which has
the correct classical limit $\cos\varphi+i\sin\varphi$ (using Eq.\ 
(\ref{quant}) and the definition of $T_+$). Again this is a creation
operator. However, it cannot be identified with $\hat{U}$ of
Subsec.~\ref{s:Restrict} as the latter operator respects the norm ---
being isometric --- while $T_+$ (or likewise $T_+H^{-1}$) does not.

The deficiency of this ansatz can be traced back to the fact that
$T_-T_+\not=H^2$, although the classical limit of this relation yields
an equality, namely $p^2\sin^2\varphi+p^2\cos^2\varphi=p^2$. This is
very similar to the difficulties of maintaining the relation
$\cos^2\varphi+\sin^2\varphi=1$ in a quantum theory of $T^*S^1$
discussed in Ref.~\cite{Isham} and we now apply a similar strategy as the
one of Isham to cure our problems here. Related issues for $\CS$ will
be discussed in detail also in the next subsection.

Classically, there are certainly various possibilities to express the
function $U=\exp(i\varphi)$ on phase space $\CS$. One such a
possibility is provided by \be \label{Uclass}
U=\frac{p\cos\varphi+ip\sin\varphi}{\sqrt{(p\sin\varphi)^2+(p\cos\varphi)^2}}
\, .  \ee This function on $\CS$ is readily translated into the
operator $T_+(T_-T_+)^{-\frac{1}{2}}$ (again using Eq.~(\ref{quant})).
Note that $T_-T_+$ is a positive, essentially self--adjoint operator
having eigenvalues $q+\lambda(\lambda+1)$ on the states
$\phi^q_{\lambda}$ so that this expression is a well--defined
operator.  (The minimal of these eigenvalues is given by $2k$
($q=k(1-k)$ and $\lambda\geq k$ for the representation $D^k$ in the
positive discrete series). Thus, the operator
$T_+(T_-T_+)^{-\frac{1}{2}}$ is well--defined only for $k>0$, which is
consistent with the fact that only under this condition the
representation $D^k$ is {\em unitary}.)  Using the Hilbert space
isomorphism between $\CH_{\theta + m_{min}}$ and $\CH_k(\CD)$ it is
then easily verified (using Eq.\ (\ref{creation})) that $\hat{U}$ and
$T_+(T_-T_+)^{-\frac{1}{2}}$ act identically on the Hilbert space and
thus may be identified.  (There are factor ordering problems in
defining $T_+(T_-T_+)^{-\frac{1}{2}}$ as a quantization of the
classical expression (\ref{Uclass}). They are, however, fixed by
asking for a quantization which acts isometrically to make possible an
identification with $\hat{U}$.)

In particular, now the adjoint of $T_+(T_-T_+)^{-\frac{1}{2}}$ is
$(T_-T_+)^{-\frac{1}{2}}T_-$, and we have the relations
$$
  \hat{U}^*\hat{U}\phi^q_{\lambda}=
  (T_-T_+)^{-\frac{1}{2}}T_-T_+(T_-T_+)^{-\frac{1}{2}}\phi^q_{\lambda}=
  \phi^q_{\lambda}
$$
for all $\lambda$, and
$$
  \hat{U}\hat{U}^*\phi^q_{\lambda}
  =T_+(T_-T_+)^{-\frac{1}{2}}(T_-T_+)^{-\frac{1}{2}}T_-\phi^q_{\lambda}=
  (1-\delta_{\lambda\lambda_0})\phi^q_{\lambda}.
$$
This demonstrates the
already known isometry and nonunitarity of
$T_+(T_-T_+)^{-\frac{1}{2}}=\hat{U}$.

Up to now we expressed the operator $\hat{U}$ obtained in
Subsection~\ref{s:Restrict} in terms of $T_+$ and $T_-$. Conversely,
we can express $T_+$ and $T_-$ in terms of $\hat{p}$ and $\hat{U}$
(cf Eq.\ (\ref{actiongms})): \be T_+ = -\hbar^{-1} \,
\sqrt{(\hat{p}+ (k-1)\hbar) \, (\hat{p}-k\hbar)} \, \widehat{U}
\label{T+} \, , \ee while $T_-=T_+^*$.

The constructions of the present subsection provide appropriate
identifications of the operators obtained in the two quantization
schemes. These results hold true also for values $k>1$, if we relax
the maximality condition when using projection quantization 
(then $m_{min}$ is not necessarily zero), which will be necessary in
Subsect.~\ref{s:Hardy} to obtain realizations of the complete positive
discrete series.

When quantizing $\CS$ by projection quantization 
we have, however, to
demand $m_{min}=0$.  If we restrict $k$ to lie in $(0,1]$ in the group
theoretical quantization, the identifications of this subsection prove
equivalence of the two approaches. In Subsec.\ \ref{s:Hardy}, we will
make this more explicit by studying the isomorphism of the respective
Hilbert spaces in terms of function spaces. 

\subsubsection{Ambiguities connected with the parameter $k$}

By comparing two quantizations, namely the group theoretical one and
the projection quantization, 
we arrived in the preceding
subsections at a one--parameter family of inequivalent quantum
theories labeled by the parameter $k\in (0,1]$. Such an ambiguity has
to be expected because of $\pi_1({\cal S})=\dZ$ (cf our discussion in
Sec.\ \ref{Standard}).

Nevertheless, one could be tempted (as, e.g., the authors of
Ref.~\cite{Louko} and Ref.~\cite{Trunk}) to restrict this
arbitrariness further by demanding that the Casimir operator
$C=T_0^2-T_1^2-T_2^2$, whose eigenvalue $q=k(1-k)$ determines a
particular representation of the positive discrete series, should be
zero (yielding $k=1$; note that $k>0$ for unitary representations).
The apparently best argument for this step would be provided by the
fact that the classical limit of $C$,
$p^2-p^2(\sin^2\varphi+\cos^2\varphi)$, vanishes identically.
However, this reasoning is not compelling: Using the group theoretical
quantization, we know the quantum operators corresponding to the
generators $p$, $p\sin\varphi$, and $p\cos\varphi$, but we cannot
unambigously determine the quantization of, e.g., $\sin\varphi$ or
$\cos\varphi$ (we have to divide by $\hat{p}$ in some appropriate
sense), the sum of whose squares was used as one in the above
conclusion. Because of factor ordering ambiguities we have to
distinguish between the operators $\widehat{p}\;\widehat{\sin\varphi}$
and $(\widehat{p\sin\varphi})$, for instance, whereas in the classical
expression we can simply factor out $p$.

Imposing $C=0$ to exclude representations with $k\not=1$ is basically
the argument provided in Ref.~\cite{Louko} (leading to Eq.\ (3.14) of
Ref.~\cite{Louko}). Also in the algebraic quantization, mainly used in that
paper, noninteger values of $k$ excluded there arise when factor
ordering ambiguities are taken into account. The argumentation in
Sec.\ 3.2 of Ref.~\cite{Trunk}, on the other hand, would even lead to the
trivial representation ({\em all\/} $T$s vanishing) as the only
quantum realization of $\CS$ (not to $k=1$ as concluded there).

As discussed above, the condition $C=0$ just imposes the relation
$(\widehat{p\sin\varphi})^2+(\widehat{p\cos\varphi})^2=\hat{p}^2$. However,
because of factor ordering ambiguities, this says nothing about the
quantum version of $\sin^2\varphi+\cos^2\varphi=1$, which would be
used as an argument for imposing it. We thus do not find it convincing
to impose $C=0$ as a condition for singling out the value $k=1$.

To round off the above discussion, we provide natural quantizations of
$\sin\varphi$ and $\cos\varphi$ inspired by the quantization of
$U=\exp i\varphi$ in the preceding subsections. As demonstrated there,
it is possible to restrict the freedom in defining
$\widehat{\sin\varphi}$ and $\widehat{\cos\varphi}$ by demanding that
the quantization of $U\equiv \cos\varphi+i\sin\varphi$ acts
isometrically. This leads to\footnote{After completing this work we
  became aware of the fact that a similar strategy has been followed
  in Ref.~\cite{Suss} (for $m_{min}=\theta=0$ in our notation,
  i.e., for parameters where the identification with the group
  theoretical quantization breaks down) in the context of quantum
  optics, where the phase space $\CS$ plays a major role (cf
  Ref.~\cite{Lynch}, we thank H.\ Kastrup for this remark).} 
\begin{eqnarray}
  \widehat{\sin\varphi} & := & -\,\frac{i}{2}\,(\hat{U}-\hat{U}^*)=
     -\,\frac{i}{2}\left(T_+(T_-T_+)^{-\frac{1}{2}}-
    (T_-T_+)^{-\frac{1}{2}}T_-\right)\label{sin}\, ,\\
  \widehat{\cos\varphi} & := & \quad\frac{1}{2}\,(\hat{U}+\hat{U}^*)=
    \quad\frac{1}{2}\left(T_+(T_-T_+)^{-\frac{1}{2}}+
    (T_-T_+)^{-\frac{1}{2}}T_-\right)\label{cos} \, ,
\end{eqnarray}
which are self--adjoint operators with the correct classical limits.
Although these expressions may appear rather complicated (as compared
to $T_1$ and $T_2$ for $p\sin\varphi$ and $p\cos\varphi$), they are
seen to come as close to the classical properties of $\sin\varphi$ and
$\cos\varphi$ as possible in the present context:
First, they satisfy
$$
  (\widehat{\sin\varphi})^2+(\widehat{\cos\varphi})^2=
  1-\frac{1}{2}P_{\lambda_0},
$$
violating $\sin^2\varphi+\cos^2\varphi=1$ only in the ground state
characterized by $\lambda=\lambda_0$ ($P_{\lambda_0}$ denotes the
projector on that state). Second, also the commutator
$$
  [\widehat{\sin\varphi},\widehat{\cos\varphi}]=\frac{i}{2}P_{\lambda_0}
$$
is nonvanishing only in the lowest state, whereas the commutators
\be 
  [H,\widehat{\sin\varphi}]=-i\,\widehat{\cos\varphi}\quad,\qquad
  [H,\widehat{\cos\varphi}]=i\,\widehat{\sin\varphi} \label{Hui}
\ee
represent the classical Poisson relations exactly.

These are only minor violations of the classical identities, which
are, moreover, independent of the value of $k\in (0,1]$. Note also
that there can be {\em no\/} self--adjoint and commuting operators $s$
and $c$ with $[H,s]=-ic$, $[H,c]=is$ which also satisfy $s^2+c^2=1$ in a
quantum theory of $\CS$.  Otherwise, the operator $c+is$ would be a
quantization of $U=\exp i\varphi$ as a {\em unitary\/} operator
generating translations, which is a contradiction according to the
discussion in Subsec.\ \ref{s:Restrict}.

\subsubsection{Different realizations of the positive discrete series\\ on
function spaces over $S^1$}
\label{s:Hardy}

By choosing the $\varphi$--representation of the Hilbert space
$\CH_{\theta+m_{min}}$ in Subsections~\ref{s:Restrict} and
\ref{s:Equiv}, we are implicitely provided with a realization of the
representation $D^k$ ($k \equiv \theta + m_{min}$) on a space of
sections of a (trivial) bundle over $S^1$ with a connection
characterized by $\theta$.  On the other hand, by restricting the
elements of the representation space $\CH_k(\CD)$ of
Subsection~\ref{s:GroupQuant} to its boundary values, we obtain a
realization of $D^k$ on a space of functions on $S^1$, too. (Similar
transitions between different Hilbert spaces have been discussed in
more detail in Ref.~\cite{Schramm}.)  Now we want to compare these two
different realizations.  In order to cover all the inequivalent
representations in the positive discrete series ($k\in\dR^+$), we drop
here the condition $k\in (0,1]$ (``physical'' representations in the
group theoretical quantization or, respectively, $m_{min}=0$
(maximality in the projection quantization)). 

To allow a comparison, we first transform the space
$\CH_{\theta+m_{min}}$ into a function space over $S^1$ as well (more
precisely, we trivialize the bundle, transferring the
$\theta$--dependence of the transition function into the momentum
operator, cf our discussion in Sec.\ \ref{Standard}).  This is done
most easily by multiplying the elements $f_{\theta,m}$ by $\exp (- i k
\varphi)$, yielding $\exp (in\varphi)$, $n \equiv m - m_{min} \, \in
\dN_0$ as the new orthonormal basis elements of a Hilbert space, which
is denoted by $H^2_+$: It is the Hardy space of the unit circle (cf
Ref.~\cite{math} for further details on this space).  Note that the
inner product is unaltered by the above transition and still provided
by $(2\pi)^{-1} \, \int d\varphi \, \overline{\psi_1(\phi)} \,
\psi_2(\phi) =: (\psi_1,\psi_2)_+$.

On $H^2_+$ the $so(1,2)$--generators are then easily seen to take
the form (cf Eq.\ (\ref{T+})): \be T_0 = \frac{d}{d\phi} + ik \; , \quad T_+
= - \exp(i\phi) \, \sqrt{\left(2k-i\frac{d}{d\phi}\right) \, 
\left(1-i\frac{d}{d\phi}\right)}  \; , \quad
T_- = T_+^* \, . \label{gen1} \ee By exponentiation this provides a unitary
(irreducible) representation of the universal covering group of
$SO^{\uparrow}(1,2)$, being a concrete realization of $D^k$ on $H^2_+$.

In the following this realization shall be compared to the one
obtained by restricting elements of $\CH_k(\CD)$ to their boundary
values on $S^1=\partial\CD$, which leads to a Hilbert space
$\CH_k(S^1)$.  Because an antiholomorphic function on $\CD$ is already
determined by its boundary values, the inner product of $\CH_k(S^1)$
is defined by anti--analytically continuing two given functions on
$S^1$ into $\CD$ and using the inner product $(\cdot,\cdot)_k$ of
$\CH_k(\CD)$ defined in Subsection~\ref{s:GroupQuant}. In this way an
orthonormal basis in $\CH_k(S^1)$ is seen to be provided by (cf Eq.\ 
(\ref{ONbasis})) \be \widetilde{g}_{k,n}(\varphi)=
\sqrt{\frac{\Gamma(2k+n)}{\Gamma(2k)\Gamma(n+1)}}e^{in\varphi}\quad,\quad
n\in\dN_0 \, . \label{ortho} \ee Here we used the coordinate $\varphi$
on $S^1$ in reversed orientation as compared to the standard
definition. This leads to $\zq=\exp(i\varphi)$, slightly simplifying
the following relations.

Note that except for $k=1/2$ there is {\em no}\/ representation of the
inner product $(\cdot,\cdot)_k$ of $\CH_k(S^1)$ in terms of an
integral over $S^1$ for some measure $\mu(\phi)$, i.e.\ there is no
function $\mu(\phi)$ such that $(\psi_1,\psi_2)_k = \int d\varphi
\,\mu(\phi) \, \overline{\psi_1(\phi)} \, \psi_2(\phi)$ except for $k=1/2$ (in
which case $\mu\equiv (2\pi)^{-1}$). This is seen most easily by inserting
the orthonormal set of wave functions (\ref{ortho}) into such an ansatz.
So (for $k > 1/2$) the continuation into the disc is an essential
ingredient in the definition of the inner product of $\CH_k(S^1)$ in
terms of an integral.

Alternatively, the inner product of $\CH_k(S^1)$ may be represented as
an ordinary $L^2(S^1)$--inner product with an {\em operator--valued}\/
metric $A_k$ (cf Ref.~\cite{Vilenkin} for further details):
$(\psi_1,\psi_2)_k = (\psi_1,A_k\, \psi_2)_+$.
(This observation shows that the Hilbert spaces $\CH_k(S^1)$ used here
are identical to the Hilbert spaces $H_{A_k}^2$ of Sec.~5.3 in
Ref.~\cite{Schramm}.)

In both cases $H^2_+$ and $\CH_k(S^1)$ we are regarding wave functions
of the form $\psi(\phi) = \sum_{n\ge 0} a_n \exp(in \phi)$.  However,
because the Hilbert spaces are completions in different inner products
the function spaces are different: the Hardy space $H^2_+$ consists of
all functions $\psi(\phi)$ with $\sum_{n\ge 0} |a_n|^2 < \infty$,
whereas in $\CH_k(S^1)$ the functions have to obey $\sum_{n\ge 0}
|a_n|^2 \Gamma(2k+n)^{-1}\Gamma(2k)\Gamma(n+1) < \infty$. It follows
immediately that {\em as function spaces}\/ $\CH_k(S^1) \subset H^2_+$
for $k<1/2$, $\CH_k(S^1)=H^2_+$ for $k=1/2$, and $H^2_+ \subset
\CH_k(S^1)$ for $k>1/2$.  While $H^2_+$ is a subspace (and thus also a
subset) of $L^2(S^1,d\varphi)$, $\CH_k(S^1)$ is a subset (but not a
subspace for $k<1/2$) of $L^2(S^1,d\varphi)$ for $k \le 1/2$ (and only
for $k \le 1/2$).

The action of the $so(1,2)$--generators in $\CH_k(S^1)$ is derived from
Eq.~(\ref{actHsD}) using $\zq=\exp(i\varphi)$ and $\zq
d/d\zq=-id/d\varphi$: \be T_0=\frac{d}{d\varphi}+ik\; , \quad
T_+= \exp(i\varphi) \left(-2k+i\frac{d}{d\phi}\right)
\; , \quad T_- = T_+^* \, , \label{gen2}
\ee where now the adjoint is to be taken with respect to the inner
product in $\CH_k(S^1)$, certainly.

Clearly, this presentation of the $so(1,2)$--generators as operators
on wave functions over $S^1$ is {\em different}\/ from the one
obtained before in Eq.\ (\ref{gen1}), {\em except}\/ for $k=1/2$ where
also $\CH_k(S^1) = H^2_+$. Eq.\ (\ref{gen1}) constitutes, to the best
of our knowledge, a novel realization (which is similar to the
Holstein--Primakoff representation of $SU(1,1)$ \cite{Gerry}) 
of the positive discrete series
on a space of wave functions over $S^1$ (namely the Hardy space).  In
the standard realization on wave functions over $S^1$, the operators
have a rather simple action (provided by Eq.\ (\ref{gen2})); however,
the corresponding, $k$--dependent Hilbert space $\CH_k(S^1)$ carries a
rather complicated and $k$--dependent inner product (cf the
discussion above). In contrast, in the other realization the Hilbert
space is simply a Hardy space with standard $L^2$--inner product,
independently of the value of $k$. The price to be paid for this
simplification of the Hilbert space is the appearance of roots of
differential operators in the representation of the
$so(1,2)$--generators (cf Eq.\ (\ref{gen1})).

Note that despite the ($k$--dependent) subset relations between
$H^2_+$ and $\CH_k(S^1)$, the $so(1,2)$--representation is certainly
still irreducible in each of the respective Hilbert spaces (as is
obvious from the Hilbert space isomorphism of Subsection~\ref{s:Equiv}); the
difference in the spaces is compensated by the different action of the
group generators.

\section{Discussion}

In Sec.\ \ref{s:quant} of this article we first motivated and recalled
the basic rules for group theoretical quantization as outlined in
Ref.~\cite{Isham}.  At the example of $T^*S^1$ it became obvious that
the strong generating principle (SGP) is an essential property to be
fulfilled by the fundamental observables of the group action.
Otherwise apparently admissible group actions can be provided which,
however, were seen to yield an unacceptable spectrum of the momentum
operator.

Checking the SGP requires the study of completeness properties of the
fundamental observables generating the group action, which may be a
cumbersome task for more involved phase spaces. Here Lemma 1 may be of
assistance: Triviality of the center of the effectively acting
projection of the canonical group was found as a necessary condition
for the validity of the SGP in a wide range of cases. 

We then pointed out that the lift of the diffeomorphism group of a
manifold $Q$ to $\CP = T^*Q$ has a transitive action on (the connected
parts of) $\CP_*$ which results from $\CP$ upon removal of the points
of vanishing canonical momenta. Since, by construction, this action
is also effective and Hamiltonian with momentum map (cf Lemma 2),
finite--dimensional subgroups of $\Diff(Q)$ are good candidates for
the use in a group theoretical quantization of such subbundles. This
strategy was applied in Sec.\ \ref{s:S} to construct the
$SO^\uparrow(1,2)$--action on $\CS = T^*S^1|_{p>0}$ as the lift of the
respective diffeomorphism group of $S^1$. Other effective actions of
covering groups of $SO^\uparrow(1,2)$, found in this way as well,
could be excluded by the SGP (cf also Lemma 1). In an appropriate
sense (cf Theorem 3) the $SO^\uparrow(1,2)$--action on $\CS$ was found
to be the unique admissible group action for quantization of the phase space
$\CS$.

In Subsec.\ \ref{s:restriction} we proposed a projection method for
quantizing phase spaces which are appropriate submanifolds of phase
spaces with known quantum realization. Examples for such submanifolds
are $T^*\dR^+$ and $\CS$: The quantum theory for the phase space
$T^*\dR^+$ ($\CS$) is obtained from the standard quantum theory of
$T^*\dR$ ($T^*S^1$) with its Hilbert space $\widetilde \CH$ upon
restriction to the maximal subspace $\CH$ on which the operator
inequality $\hat q > 0$ ($\hat p > 0$) is satisfied. The corresponding
(unique) projection operator from $\widetilde \CH$ to $\CH$ may then
be used also to obtain operators defined originally only within
$\widetilde \CH$.

We outlined some of the prerequisites for the applicability of the
projection method of quantization as well as its basic rules. It may
well be that the study of further examples will lead to an adjustment
and refinement of these ideas. A promising strategy is also to employ
the technique of symplectic cuts \cite{scuts} for a comparison of the
projection quantization with more standard quantization schemes, which is
done in Ref.\ \cite{scutproj}. 

As a possible arena for its application we discussed issues in quantum
gravity, where nondegeneracy of the metric has to be imposed. In this
context we remarked that the presence of constraints may lead to
subtleties (in addition to the well--known problems of the
quantization of constrained systems).

For $T^*\dR^+$ the quantum theory resulting from projection quantization 
is equivalent to the standard one for this phase space. For $\CS$ it
coincides with the quantum theory obtained from group theoretical
quantization, if in addition to the negative discrete and the
continuous series of the $so(1,2)$--representations (cf.\ Subsec.\ 
\ref{s:so} and Ref.~\cite{Schramm}) also representations of the
positive discrete series $D^k$ with $k >1$ are discarded.  {\em
  Within\/} the (present--day) scheme of group theoretical
quantization this may be justified only by declaring them to be
unphysical representations. In lack of a truely physical realization
of the phase space $\CS$, the above ``unphysical'' is not to be taken
too literally. However, the resulting restriction, leaving only $D^k$
for $0< k \le 1$ as possible Hilbert spaces for $\CS$, agrees also
with what one would expect on general grounds in the context of
geometric quantization.
Projection quantization 
yields a one--to--one relation between the
$\theta$--angle of the quantum theories of $\CS$ and $T^*S^1$.  Thus,
agreement with other approaches to the quantization of $\CS$ forces us
to truncate the range of allowed values of $k$ in the group
theoretical one.

Within this paper we always tried to keep track of possible
ambiguities in the transition from the classical to the quantum
system. In the group theoretical approach this lead us to always
consider representations of the universal covering of the group with
admissible action. Note, however, that in all the examples studied
the fundamental group of the phase space was at most $\dZ$. The
situation may become more involved for the case of nonabelian
fundamental groups \cite{Nico}. 

\section*{Acknowledgement}
We are grateful to H.\ Kastrup for several critical remarks and
questions which helped improving the material of this paper. We are
also particularly indebted to D.\ Giulini for a critical reading of
the manuscript and his remarks. We finally mention fruitful
discussions with A.\ Alekseev, N.\ D\"uchting and F.\ Schramm.

This work was concluded during a visit of the second author at the Erwin
Schr\"odinger Institut (ESI) in Vienna; the support of ESI is kindly
acknowledged.

\end{document}